\documentclass[pdflatex,sn-apa]{sn-jnl}

\usepackage{graphicx, makecell, array, tabularx, multirow, hhline, longtable, booktabs}%
\usepackage{amsmath,amssymb,amsfonts,bm,setspace}%
\usepackage{amsthm}%
\usepackage{mathrsfs}%
\usepackage[title]{appendix}%
\usepackage{xcolor}%
\usepackage{textcomp}%
\usepackage{manyfoot}%
\usepackage{listings}%
\usepackage{afterpage, pdflscape}%
\usepackage{rotating}%
\usepackage{comment}

\graphicspath{ {./pics/} }

\title[A Dynamical Cartography of the Epistemic Diffusion of Artificial Intelligence in Neuroscience]{A Dynamical Cartography of the Epistemic Diffusion of Artificial Intelligence in Neuroscience}

\author*[1]{\fnm{Sylvain} \sur{Fontaine}}\email{sylvain.fontaine@cnrs.fr}

\affil[1]{\orgdiv{GEMASS}, \orgname{CNRS-Sorbonne Universit\'e}, \orgaddress{\street{59-61 rue Pouchet}, \postcode{75017} \city{Paris}, \country{France}}}

\abstract{
Neuroscience and AI have an intertwined history, largely relayed in the literature of both fields.
In recent years, due to the engineering orientations of AI research and the monopoly of industry for its large-scale applications, the mutual expansion of neuroscience and AI in fundamental research seems challenged.
In this paper, we bring some empirical evidences that, on the contrary, AI and neuroscience are continuing to grow together, but with a pronounced interest in the fields of study related to neurodegenerative diseases since the 1990s.
With a temporal knowledge cartography of neuroscience drawn with advanced document embedding techniques, we draw the dynamical shaping of the domain since the 1970s and identified the conceptual articulation of AI with this particular subfield mentioned before.
However, a further analysis of the underlying citation network of the studied corpus shows that the produced AI technologies remain confined in the different subfields and are not transferred from one subfield to another.
This invites us to discuss the \textit{genericity} capability of AI in the context of an intradisciplinary development, especially in the diffusion of its associated \textit{metrology}.
}


\begin{document}

\maketitle

\section{Introduction}

In recent years, Artificial Intelligence (AI) arises to be a set of knowledge, and especially technologies, that automate and accelerate a variety of complex tasks originally performed by humans.
Supported by recent research programs and strong public-private partnerships, AI has become an essential part of fundamental research in all disciplines \citep{gargiulo_meso-scale_2023,gao_quantifying_2023} and also in the research and development sector within specialized companies \citep{ahmed_growing_2023,frank_evolution_2019}, with proven technical applications in everyday life \citep{baruffaldi_identifying_2020,xu_artificial_2021}.
Designed in interdisciplinary environments at all levels of the production of knowledge and applications, AI is currently used in many fields of research for their own purposes \citep{gargiulo_meso-scale_2023}.

Neuroscience is not immune to the pervasive influence of AI in the scientific realm.
As one of the main originating research domains of these intelligent technologies, neuroscience -- alongside the close field of cognitive science -- has a long history with AI punctuated by fruitful advancements, the most famous example being neural network architectures \citep{macpherson_natural_2021,hassabis_neuroscience-inspired_2017}.
The results of these interactions between the two fields continue to benefit both, from the improvement of current AI architectures for technological and software developments, to applications in biomedical and clinical research through the treatment of complex datasets with various kinds of data that need to be compiled, especially in clinical neurology and in neurodegenerative diseases' studies \citep{macpherson_natural_2021,gopinath_artificial_2023}.
In the context of innovation studies, this process of co-production of AI and neuroscience knowledge depicted in the neuroscience literature (although currently challenged by the scientific communities of the two fields \citep{nmi_neuroAI_2024}) can be seen as the mutual expansion of their respective \textit{adjacent possibles}, i.e., their respective research orientations are interdependent and would not exist without each other \citep{kauffman_investigations_2000,monechi_waves_2017,bianchini_artificial_2022}.
Consequently, one field is contributing to the rise of innovations into the other.

Despite this apparent beneficial mutual feedback between these two fields, an significant literature deplores the increasing distance of AI from fundamental brain research, especially due to an engineering turn that has contributed to the industrialization of some of their respective research \citep{ahmed_growing_2023,fregnac_neuro_2017,nmi_neuroAI_2024}.
By becoming increasingly practiced inside tech companies \citep{frank_evolution_2019}, which focus on the construction and training of large-scale deep learning models that require high-performance computing infrastructures, AI research is subject to a progressive thematic closure around these developments, which are substantially oriented towards short-term technological applications \citep{ahmed_dedemo_2020,klinger_narrowing_2022}.
Some neuroscientists and cognitivists are therefore advocating for a return of AI to the heart of fundamental brain research, both for the benefit of the latter and to improve current AI technologies, without getting caught up in research that would be data-driven or focused solely on technological and large-scale industrial applications \citep{perconti_deep_2020,fregnac_neuro_2017}.

By focusing on academic neuroscience research, Fontaine et al. (\citeyear{fontaine_epistemic_2024}) also exemplified this distance between AI and neuroscience, mainly based on disciplinary and social aspects.
Indeed, they showed that the diffusion of AI across neuroscience since 1970 remains limited and primarily reserved to a dedicated workforce organized in a socio-cognitive environment confined in the neuroscience ecosystem.
This closure of AI research into a dedicated specialty in neuroscience therefore indicates that AI is not intended to replace traditional scientific reasoning methods in the field or the practical applications in clinical routines.
It thus supports recent work in sociology of science that unveils the difficulties and reluctance expressed by some scientific and medical communities to apprehending these AI tools -- which also requires a technological acculturation to them --, especially radiologists confronted with new software for automatic anomaly recognition \citep{anichini_radio_2021,mignot_radio_2022}.
Such behaviors could precisely hinder the dissemination of these technologies and knowledge across the studied social group.

All of these previous assertions invite us to conceptualize AI as a \textit{research-technology} \citep{shinn_transverse_2002,marcovich_science_2020,hentschel_periodization_2015}, i.e., a collaborative research environment involving various actors (academics, industrialists, public governance) who work together in the development of a technological instrument in which they have interest for various purposes.
Such an arena is animated by the objective of disseminating technological outputs outside itself.
These technologies are intended to be adapted to a wide range of users and be transferable to any epistemic framework, at which stage the instrument achieves a \textit{generic} character.
Fontaine et al. (\citeyear{fontaine_epistemic_2024}) investigated such a genericity process in neuroscience with bibliometric indicators, such as collaborations, which are useful to measure the intensity of social adoption, and citations to study interactions between AI's and neuroscience's disciplinary ecosystem.
Nevertheless, according to Joerges, Marcovich and Shinn, genericity is also achieved when the instrument's \textit{metrology}, here referred to as a common language and vocabulary to grasp the instrument and its actions, as well as the means by which its performances must be measured and controlled (such as the notions of \textit{benchmark}, \textit{time complexity}, \textit{representation}, \textit{training data}, in the case of \textit{machine learning}, among others), becomes pervasive in the scientific practice, reasoning and discourse of researchers in any discipline.

In this paper, we explore the long-term diffusion of AI knowledge and tools into neuroscience under this framework of genericity, in particular under the aspect of metrology diffusion and integration, by asking the following research questions:
\begin{itemize}
    \item How are AI knowledge and tools distributed across the epistemic landscape of neuroscience? In which of its knowledge subfields does AI mostly develop?
    \item How much is its conceptual universe integrated into each subfield's one?
    \item How are the AI productions within these subfields then promoted across the whole domain?
\end{itemize}

We propose to approach this process by a large-scale semantic analysis of the neuroscientific literature extracted from bibliometric databases and spanning a wide temporal period, here between 1970 and 2020 -- as main AI-related realizations emerged at the beginning of this period.
More precisely, we apply the large language model SPECTER \citep{cohan_specter_2020} on titles and abstracts of the articles included in this literature, in order to represent the whole neuroscience knowledge landscape as a two-dimensional, geographical, temporal cartography that we divide in epistemic regions -- also called \textit{subfields} in the following lines.
With such a map, we identify the regions where AI is produced and with which intensity of development.
In addition of this macro-scale analysis, we build a temporal co-occurrence network for each subfield using the concepts labeling their papers, in order to assess the degree of integration of the AI vocabulary into their respective conceptual universe.

Then, we study to which subfields such AI research is beneficial with the citation network between the articles.
Despite the various interpretations of the citation act \citep{bornmann_citations_2008}, we consider it as a core vector of knowledge transfer between two published articles.
This view is particularly prevalent in recent scientometric studies that have examined large-scale knowledge flows between several disciplines or within a unique one, at the level of either disciplinary journals \citep{petrovich_neuro-philo_2024} or concepts \citep{di-bona_decentralization_2023}.
This article proposes an alternative way of studying knowledge flows by exploiting the whole semantic information of each article in our database. 
Specifically, this approach allows to assess the semantic distance between two articles connected by a citation link on the neuroscience knowledge map: the longer the distance, the more different the research context in which the cited article has been used. 
Such a distance also captures the capability of the concepts or methods introduced in a paper to be disseminated extensively or not in the semantic space. 

In summary, we explore on the one hand the hypothesis that AI-related knowledge and technologies are more suitable to treat some neuroscience subjects than others, given the evolution of the research orientations adopted by the whole field under study,
and on the other hand the ability of AI to be transferred in different subfields of neuroscience, and therefore the degree of its \textit{genericity} in the latter caused by the potential dissemination of its own metrology.
We especially observe that AI remains confined in some local dynamics of citations in different regions of the knowledge space, thus also showing the limit of the genericity criterion at the macroscale of neuroscience as defined by Shinn et Joerges (\citeyear{shinn_transverse_2002}): the AIs designed in neuroscience is not beneficial to all subfields of neuroscience.
We finally discuss the results and propose that we are witnessing in the co-development of AI with some specific subfields of neuroscience, especially the recent studies of neurodegenerative diseases.


\section{Data and methods}
\label{sec:data}

\subsection{The dataset}
\label{sec:dataset}

\begin{figure}[htb]
    \centering
    \includegraphics[width=0.9\textwidth]{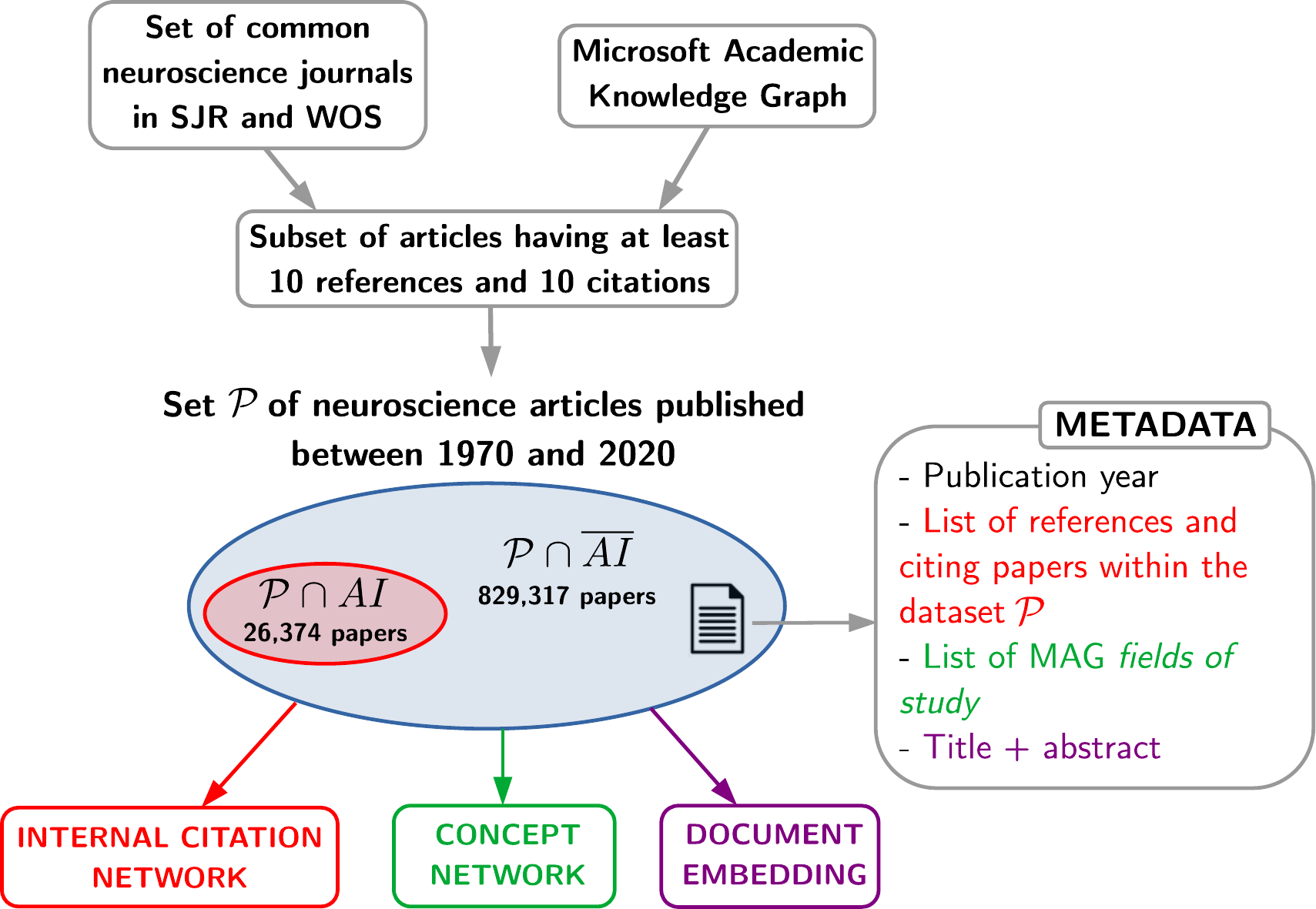}
    \caption{Building step of the dataset provided by Fontaine et al. (\citeyear{fontaine_epistemic_2024-1}) and used in this study.
    The internal citation network is defined by the citation links between the publications within the set $\mathcal{P}$.}
    \label{fig:dataset_building}
\end{figure}

This study relies on the scientometric dataset provided by Fontaine et al. (\citeyear{fontaine_epistemic_2024-1}) and uploaded in Zenodo, which was extracted from a dump file of the Microsoft Academic Graph \citep{ghidini_microsoft_2019}, shortly noted as MAG in the following lines.
This dataset has been built following the scheme in Fig.~\ref{fig:dataset_building}, which summarizes the delimitation of the neuroscience corpus and the AI-related articles inside.

In order to extract a coherent set of neuroscience publications from the MAG, Fontaine et al. (\citeyear{fontaine_epistemic_2024}) have drawn upon a list of peer-reviewed journals (no conference proceedings are considered here, and any other publication format), that are representative of the domain under study.
In particular, they combined two requests for venues labeled explicitly as “Neuroscience” for the Scimago Journal Rank (SJR) and the Web of Science (WoS), because the former has a broader and more thematic diverse coverage in neuroscience journals than the latter.
This combination, representing the union set of these two requests, results in a list of 421 journals. 
From the latter, Fontaine et al. then selected articles published between 1970 and 2020 with at least 10 bibliographical references and at least 10 citations within MAG in 2020. 
Although these thresholds were chosen both for technical reasons (reducing the size of the dataset without compromising its comprehensiveness) and for studies' objectives that differ from those of our present article, they are not detrimental to our study. 
Indeed, with regard to the first aforementioned threshold, the average number of bibliographic references per paper has been higher than 10 since 1970 \citep{lariviere_aging_2009}, so we can reasonably consider that we capture the majority of the neuroscience literature over the chosen fifty-year period according to this criterion. 
With regard to the second one, as high-impact papers in a given domain are (often) representative of its main thematic trends over a given period, we consider that focusing on neuroscience ones is suitable in the context of our study to represent faithfully the main semantic landscape of this domain over time.
The final dataset, denoted as $\mathcal{P}$ in the following and in Fig.~\ref{fig:dataset_building}, contains 855,691 papers.
Of these, 26,374 include some AI-related keywords in their respective title and/or abstract.
These AI keywords, provided by Gargiulo et al. (\citeyear{gargiulo_meso-scale_2023}) in their supplementary material, are extracted from dedicated glossaries on Google and Wikipedia, and are commonly used by experts for bibliometric studies and policy prospective for national AI research \citep{baruffaldi_identifying_2020,hajkowicz_csiro_2022}.

For the purposes of our study, we select three types of metadata associated with all the papers within this corpus $\mathcal{P}$: their titles and abstracts, their bibliographic references and citations received within $\mathcal{P}$, and their \textit{fields of study} automatically set by MAG.  
In the following subsections, we detail the construction of temporal, aggregated data structures from these metadata, which will be used to represent the neuroscience knowledge space and to study the diffusion dynamics of AI throughout this space. 

\subsection{Representing the knowledge space of neuroscience}
\label{sec:build_map}

Inspired by recent efforts to map scientific knowledge as faithfully as possible \citep{singh_charting_2024,liu_science_2024,gonzalez-marquez_landscape_2024}, we need a large language model (LLM) capable of transforming the textual metadata of the articles in our corpus, here their titles and abstracts, into vectors that are easier to manipulate numerically, representing their knowledge and the contextual features in which it is stated (through specific sentence constructions or recurrent word associations), in order to evaluate the degree of lexical similarity between two given papers.
Here, we especially use the SPECTER model \citep{cohan_specter_2020}, a BERT-based sentence transformer inherited from the SciBERT model \citep{beltagy_scibert_2019}, which has been trained on a scientific corpus mixing computer science, biology, and medical science.
In particular, SPECTER shows a better disciplinary comprehension of the scientific documents than SciBERT, since the former has been trained on a corpus where two given elements can be linked by a citation relation.
Instead of generating an embedding of a document based only on its own semantic context (\textit{intra}-document context), SPECTER also considers the semantic context of other documents related to the former by citations (\textit{inter}-document context).

\begin{figure}[t!]
    \centering
    \includegraphics[width=\textwidth]{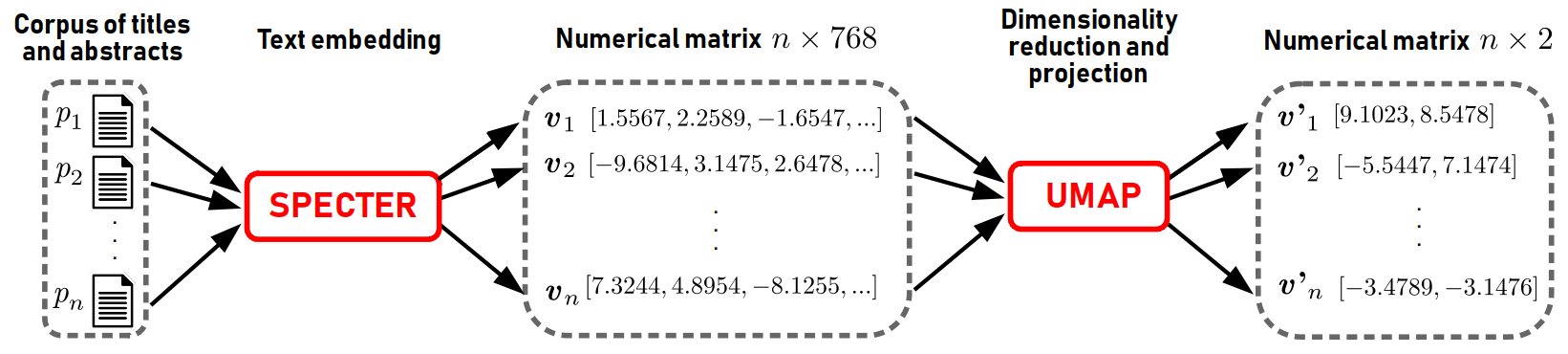}
    \caption{Schematic representation of the pipeline used to generate the neuroscience knowledge map below.
    First, SPECTER converts the textual elements of the articles within our corpus (called $p_i$) into 768-dimensional vectors each ($v_i$), then UMAP transforms the latter into 2-dimensional vectors ($v'_i$) given the data structure provided by SPECTER.
    The final list of 2D points are used to draw the map.}
    \label{fig:nlp_pipeline}
\end{figure}

In order to produce a synthetic cartography of neuroscience knowledge, we first apply SPECTER to the textual elements of our corpus, here the titles and abstracts of its papers, in order to obtain their respective contextual embedding, which result as vectors of 768 floating elements each.
We then reduce these vectors to a two-dimensional space with a \textit{Uniform Manifold Approximation and Projection} (UMAP), which aims to preserve the proximity (or similarity) between the initial vectors expressed in the higher-dimensional space, here of dimension 768 \citep{mcinnes_umap_2020}.
This lower-dimensional space is thus an approximation of the original semantic space that characterizes all of our neuroscience work.
This procedure is illustrated in Fig.~\ref{fig:nlp_pipeline}.
This step finally allows us to plot the two-dimensional points in such a space, as shown in Fig.~\ref{fig:map}.
A paper is represented by a single point.
An aggregate of such points very close to each other in a specific region of this map thus forms a vocabulary subspace where the knowledge inscribed in the associated articles is similar.
If such a subspace is isolated in the maps, the papers belonging to it thus compose a single research topic.

Since the initialization of the UMAP projection is random, and because of the stochastic property of the algorithm,\footnote{We refer the reader to: \url{https://umap-learn.readthedocs.io/en/latest/reproducibility.html}.} we remind the reader that the reduction obtained is not unique.
Nevertheless, since UMAP preserves the local proximity of the projected points, the reductions obtained after several realizations of the algorithm would have the same shape in a two-dimensional plane but oriented or organized differently, thus not affecting a further clustering process on these reduced data.
Therefore, we propose here an investigation of one realization of this projection.
App.~\ref{app:knn} details further analyses of the robustness of the UMAP embedding according to various features, such as the location of the nearest neighbors of each paper and their topical proximity -- based on the recent \texttt{Topic} classification of scientific papers given by OpenAlex.

Then, we apply a \textit{Hierarchical Density-Based Spatial Clustering of Applications with Noise}, or HDBSCAN \citep{mcinnes_hdbscan_2017}, to this set of ``reduced'' points.
This method first returns a condensed dendrogram whose sticks are clusters of points, which could be merged or divided by varying a density threshold $\lambda$.
The partitioning of the dataset is detailed in App.~\ref{app:dendogram}.
We retain nine clusters, whose names are established with the distributions of the MAG \textit{fields of study} and OpenAlex \textit{primary topics} and \textit{keywords} related to the papers within them, as detailed in App.~\ref{app:definition_cluster}.
Such a static clustering facilitates the comparison of the different development stages of every cluster period by period, and the comparison of the development of AI within them and its diffusion capability across the knowledge space with respect to its originating epistemic area.

Finally, we represent the reduced dataset as a density plot in Fig.~\ref{fig:map}, with clusters delineated by solid colored lines.
We draw on the same dataset and partition to examine a series of temporal snapshots of this map, shown in Fig.~\ref{fig:map_temporal}.

\subsection{Assessing the centrality of AI-related concepts in each cluster's concept network}
\label{sec:build_concept_net}

As suggested in Introduction, the diffusion and the integration of a technology in everyday scientific practice, e.g., AI, is conditioned by the dissemination of related concepts and vocabulary, which could be approached with the temporal concept network describing each knowledge cluster of the map built above.
To build such a concept network, we use the \textit{fields of study},\footnote{Each field of study associated with one given paper is also scored with the probability of attaching this paper to these field of study. Here we don't use this score because we consider that the appearance of one field of study for one paper is sufficient to assume that the latter belongs to the former.} also called \textit{concepts} throughout this article, which are thematically labeling all the papers in our dataset $\mathcal{P}$.
These fields of study are also organized in an arborescence by MAG (namely a directed acyclic tree) and hierarchically classified by \textit{level}.
This parameter indicates the level of specificity of a given field of study in the science system, from very general one such as discipline or sub-discipline levels, i.e., levels 0 and 1 respectively, to research subjects or keywords extracted from abstracts, i.e., levels 4 and 5 respectively.
Such fields of study are thus proxies for delimiting big or small knowledge domains.
In what follows, for the sake of consistency and to provide a good balance between these different levels of specificity, we consider only the fields of study situated at levels 2 or 3.
A robustness check performed in App.~\ref{app:conceptNet_topo} supports this choice, which preserves at most the main topological structures of the original, all-levels concept network to other choices of level ranges.
With such a selection, we cover 93\% of the papers in $\mathcal{P}$ labeled with at least two concepts at these levels, and 96.5\% of the AI-related papers in the subset $\mathcal{P}\cap AI$.
From the entire MAG concept arborescence, we define AI-related concepts as those located at levels lower than 1 and inherited from either the concepts \textit{Artificial intelligence} or \textit{Machine learning}, which are both at level 1.

\begin{figure}[t!]
    \centering
    \includegraphics[width=0.9\textwidth]{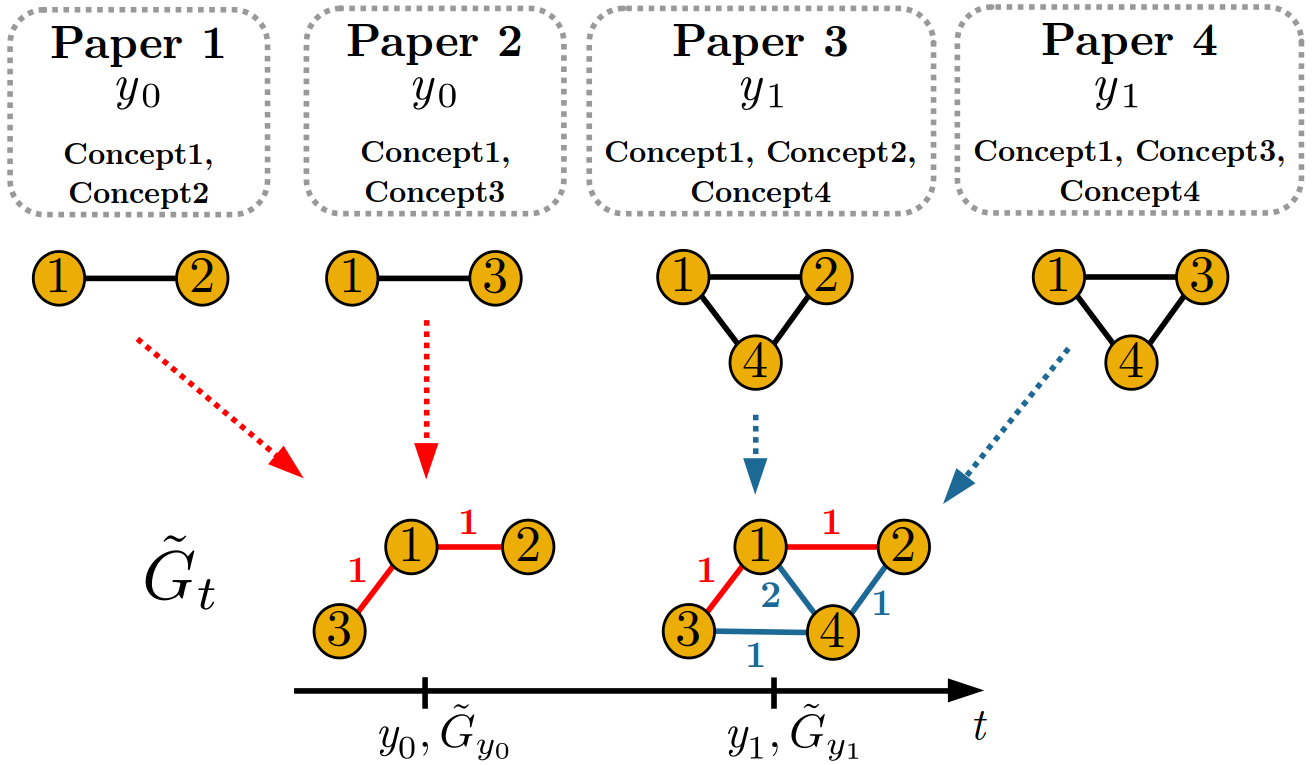}
    \caption{Schematic representation of the building of the temporal concept network, denoted as $\tilde{G}_t$ as in the main text, based on four articles published in years $y_0$ or $y_1$. The weights of the edges are indicated by the numbers next to them.}
    \label{fig:conceptNet_building}
\end{figure}

In this section, we build a dynamical indicator to assess the degree of entanglement of AI-related fields of study within the concept network of a given cluster, in particular by determining their location within such a network -- in the core or in the periphery.
The concept network describing the knowledge universe of a given cluster is framed under the cumulative process as follows and shown in Fig.~\ref{fig:conceptNet_building}.
Starting from the co-occurrence network of level-2 and/or level-3 fields of study present in the cluster's papers published in 1970, denoted as $G_0=(V_0,E_0)$ -- where $V_0$ the set of fields of study and $E_0$ the set of co-occurrence edges between the latter -- and the next one embodied in the papers published in 1971, denoted as $G_1$, we construct the cumulative concept network for the year 1971 as the network $\tilde{G}_1=G_0 \cup G_1 = (\tilde{V_1},\tilde{E_1})$, where $\tilde{V}_1=V_0 \cup V_1$ and $\tilde{E}_1=E_0 \cup E_1$. 
We generalize this recursive relation to any year $t$ as $\tilde{G}_t = G_{t-1} \cup G_t = (V_{t-1} \cup V_t,\, E_{t-1} \cup E_t)$, so that the final graph $\tilde{G}_{2020}$ is the entire concept network of the cluster spanning the period 1970-2020.
This temporal dynamic of concept co-occurrence thus gives rise to an evolving semantic network revealing the topical evolution of the cluster under study, where the meaning of any concept depends on its neighboring ones, with which its links may be added or strengthened over time \citep{rule_lexical_2015,cheng_how_2023}.

This cumulative representation of associated fields of study within the clusters allows us to track the location of AI-related ones as new concepts and links are added to the network describing the cluster over time, i.e. as the cluster's knowledge universe expands.
For example, for an AI-related concept situated in the core of the network in a given year -- which is a signal of close conceptual ties with its neighbors, and therefore of a semantic proximity to them -- do we observe a perpetuation of its position through the reinforcement of existing links in subsequent years, or its retreat toward the periphery?
We could also ask the opposite in the case of a peripheral AI concept advancing toward the core as the network grows, which would be a signal not only of its increasing use but also of its ability to be grafted onto concepts of different kinds.

We define this location of concepts within the studied network by their core number obtained by a $k$-core decomposition of the network, i.e. their membership to a subgraph with a degree that is at most $c$ (the $c$-shell) but not in the next core with minimal degree $c+1$ \citep{batagelj_om_2003}.
We normalize this metric by the maximum core number of the network, i.e. $\tilde{c}=c/c_{max}$, such that the closer to 0 the indicator $\tilde{c}$, the farther away from the core the concept is -- and \textit{vice versa} when moving closer to 1.
We will refer to this indicator as \textit{coreness} in the following, and we will focus especially on the values exhibited by the AI-related fields of study within each cluster of the neuroscience knowledge map.

\subsection{Measuring the diffusion of AI-related work in the knowledge map}
\label{sec:map_AI_diffusion}

\subsubsection{Building of the citation network}

We propose to analyze the diffusion of AI-related knowledge on the neuroscience cartography drawn above through the underlying citation network between the mapped articles from the set $\mathcal{P}$.
This network is composed of 13,802,379 edges between 836,222 articles -- thus representing 98\% of $\mathcal{P}$, the remaining 2\% being not linked to any other nodes in the network.

We first define $G=(V,E)$ the citation network between all these papers, such that $V=\lbrace p_i \rbrace_{i=1,...,N}$ is the set of papers, which are identified by their respective cluster $C_k$ ($k=0,...,8$), and $E=\lbrace (p_i,p_j) \rbrace_{i,j\in [1,...,N]^2}$ the set of unique directed links between the papers in $V$, e.g. a link $(p_i,p_j)$ denoting that $p_i$ cites $p_j$.

\subsubsection{Individual citation radius of gyration of the papers}
\label{sec:gyradius_calculation}

Here, we propose a measure of the ability of a paper to influence a wide range of the neuroscience knowledge space, based on its individual radius of gyration (RoG) generated by other articles citing it -- observed in 2020.
This quantity is expressed as the square root of the moment of inertia associated with the publications citing a focal paper $i$:
\begin{equation}
r^i_g=\sqrt{\frac{1}{N_i}\sum_{j\in p_i}d_{ij}^2}\, ,
\label{eq:gyradius}
\end{equation}
where $p_i$ is the set of papers citing $i$ (with cardinality $N_i$) and $d_{ij}$ is the spatial distance between papers $i$ and $j$ in the knowledge map, i.e. the paper $i$ is considered as the citation \textit{center of mass}.
This RoG is also useful for assessing the diffusion or concentration of citations in a local vocabulary space or beyond, and with what intensity.
To compute the RoG on a polygon in the knowledge map, which must be at least a triangle, we draw our analysis on papers that have received at least 3 citations from our corpus $\mathcal{P}$.
Under this filter, we cover 639,915 papers, which account for 75\% of all papers in $\mathcal{P}$, and which concentrate 99\% of all the citation links of the network built in the previous subsection.

However, an article located at one of the edges of the map -- top, bottom, left, or right extremities -- may have a larger RoG than another article situated rather at the center of that map.
Indeed, as a result of the construction of the map, the articles near the center are situated in a middle mass of knowledge that is almost equidistant from all others, which allows their knowledge to be disseminated more easily across the entire map than that from articles located close to its borders.
For example, in Fig.~\ref{fig:map}, a paper in the cluster \textit{Parkinson's disease} ($C_4$) that cites another in \textit{Foundations of connectionist AI} ($C_1$) must have spanned a larger lexical gap than another paper also citing $C_1$ but located in the neighboring clusters on brain neural network and neurological disorders ($C_7$).

To compare the RoG between the clusters, we normalize the observed individual RoG of a given article by another hypothetical one, its \textit{maximum RoG}, which is a function of the maximal distance between this article and the one farthest away from it on the map, indicated by $d_i^{max}$ below. 
By considering the same number of papers citing the article $i$, but all hypothetically located at a distance $d_i^{max}$, the application of Eq.~\ref{eq:gyradius} gives:
\begin{equation}
r^i_{g,max}=\sqrt{\frac{1}{N_i}\sum_{j\in p_i}\left(d_i^{max}\right)^2}=\sqrt{\frac{1}{N_i}\times N_i\left(d_i^{max}\right)^2}=d_i^{max}\,.
\end{equation}
This maximum RoG is thus analogous to the maximum knowledge area that an article could cover if cited by these very distant papers on the map.

Then, we derive this alternative RoG used throughout this paper, which is actually the intensity of the observed citation spread compared to its own possible maximal one in the neuroscience map:
\begin{equation}
\tilde{r}^i_g=\frac{r^i_g}{r^i_{g,max}}\,.
\label{eq:alternative_gyradius}
\end{equation}
Since its values are comprised between 0 and 1, the closer the new RoG is to 1, the wider the diffusion, and conversely, the closer it is to 0, the narrower the diffusion.

In what follows, we will refer to this metric to compare the respective spreading ability of AI-related ($\mathcal{P}\cap AI$) and non-AI ($\mathcal{P}\cap \overline{AI}$) papers across the knowledge landscape of neuroscience.

\section{Results}
\label{sec:results}

\subsection{Where is AI situated in the knowledge landscape of neuroscience?}
\label{sec:knowledge_map_study}

\subsubsection{The topical space of neuroscience}
\label{sec:map}

\begin{figure}[t!]
    \centering
    \includegraphics[width=\textwidth]{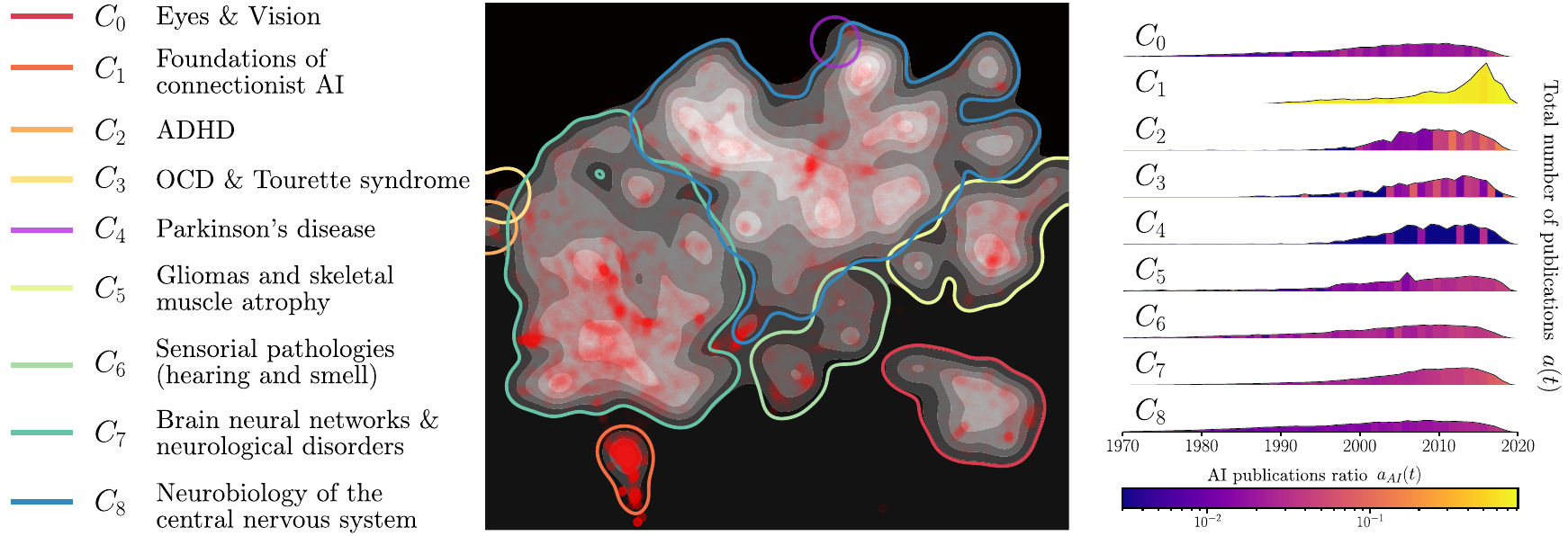}
    \caption{Neuroscience knowledge map.
    In the center is the time-aggregated knowledge map of neuroscience between 1970 and 2020.
    The grayscale represents the density of articles in this space: the lighter the zone, the denser the local concentration of papers, and vice versa when moving towards through darker shades of gray.
    The clusters delimited by colored lines, whose names are given in the legend on the left of this figure, have been drawn with the largest level line obtained with the HDBSCAN algorithm.
    The red dots are the AI-related papers distributed in this knowledge space.
    On the right are plotted the temporal evolution of the size of the clusters, normalized by their own size in the time-aggregated map on the left.
    The vertical axes of all plots are standardized and do not exceed 0.15.
    The colored area under these curves represents the temporal evolution of the share of AI-related publications within these clusters.}
    \label{fig:map}
\end{figure}

Fig.~\ref{fig:map} shows the cartography of all the knowledge encoded in the textual metadata of articles in our neuroscience database spanning the period 1970-2020, built with the method exposed in Sect.~\ref{sec:build_map}.
More precisely, this map represents the spatial distribution of these articles according to their own vocabulary, represented here as points in a broad lexical space.
For example, the vocabulary employed in the cluster \textit{Eyes \& Vision} is quite distinct from that used in the cluster \textit{Mathematics for Connectionist AI}, both of which are also different from the vocabulary employed in \textit{Parkinson's disease} area.
The names of the different subspaces delimited by colored lines -- the clusters -- are derived from the most frequent concepts associated with the papers in them, which are provided by three topical classifications, namely the MAG fields of study, and OpenAlex's topics and keywords (see App.~\ref{app:definition_cluster} for their definitions).
They are also listed in Tab.~\ref{tab:share_AI_clusters}.
We used grayscale density level lines to highlight the heterogeneous density of papers in this knowledge space.
Only AI-related publications are plotted as red dots on this map.

The spatial distribution of AI papers in this map shows at first glance that AI is able to integrate itself into different knowledge subspaces of neuroscience and also to link to a variety of knowledge across the discipline, thus again demonstrating the adaptation of AI to neuroscience epistemic objectives and a certain genericity that could also be denoted as epistemic.

However, the temporal evolution of the number of AI publications in each cluster, shown in the right-hand plots of Fig.~\ref{fig:map}, along with the distributions of AI publications among the clusters summarized in Tab.~\ref{tab:share_AI_clusters}, both testify to the uneven distribution of AI papers among the clusters.
For example, the cluster on foundations of connectionist AI ($C_1$) condenses 22.5\% of all AI publications into a much smaller knowledge space than that covered by central nervous system studies ($C_8$), which nonetheless exhibits an equivalent share of such publications (24.1\%).
The first cluster thus shows a very high density of AI articles with very similar vocabularies, while such publications belonging to the second cluster are more spread out in its broad lexical space - with some small redder subspaces of high local density, however.
Moreover, according to Tab.~\ref{tab:share_AI_clusters}, the leftmost clusters on the map, grouping formal studies in foundations of contemporary, connectionist AI-related models ($C_1$), attention-deficit and hyperactivity disorders (ADHD, $C_2$), obsessive-compulsive disorders (OCD, $C_3$) and brain neural networks and neurological disorders ($C_7$), aggregate 59\% of all the AI-related papers, thus demonstrating that the core of such publications is situated in a particular knowledge subspace of neuroscience.
The first one, although separated from the continuum of papers ranging from neurological disorders studies to neurobiological studies of the central nervous system (represented by $C_8$), remains quite close to the former, with some bridges between them through other AI publications.
This suggests that the subfield gathering the studies of the mathematical and computational foundations of connectionist AI, which could be related to computational neuroscience, preferentially maintains some links with the subfield of neurological damage studies.

\begin{table}[t!]
    \centering
    \begin{tabular}{|c|c|c|}
        \hline
        \multirow{2}{*}{\textbf{Cluster}} & \multirow{2}{*}{\textbf{Name}} & \textbf{Share of} \\
        & & \textbf{AI papers (\%)}\\
        \hhline{|=|=|=|}
        $C_0$ & Eyes and vision studies & 3,7 \\
        \hline
        \multirow{2}{*}{$C_1$} & Mathematical and computational & \multirow{2}{*}{22,5} \\
        & foundations of connectionist AI &  \\
        \hline
        $C_2$ & Attention-Deficit/Hyperactivity Disorders & 0,33 \\
        \hline
        \multirow{2}{*}{$C_3$} & Obsessive-Compulsive Disorders and & \multirow{2}{*}{0,14} \\    
         & Tourette syndrome studies & \\
         \hline
        $C_4$ & Parkinson's disease & 0,02 \\
        \hline
        $C_5$ & Gliomas and skeletal muscle atrophy studies & 5,76 \\
        \hline
        $C_6$ & Studies of sensorial pathologies & 3,02 \\
        \hline
        \multirow{2}{*}{$C_7$} & Brain neural networks and neurological & \multirow{2}{*}{36,1} \\
         &  disorders studies (epilepsy and schizophrenia) & \\
        \hline
        \multirow{2}{*}{$C_8$} & Studies of neurobiological mechanisms & \multirow{2}{*}{24,1} \\
        & in the central nervous system &  \\
        \hline
    \end{tabular}
    \caption{Distribution of AI-related articles across the clusters of the knowledge map. 
    For example, $C_0$ owns 3.7\% of all the AI publications in our dataset.}
    \label{tab:share_AI_clusters}
\end{table}

\subsubsection{Co-development of AI within neuroscience through the rise of the neurological disorders studies}
\label{sec:co-development}

\begin{figure}[t!]
    \includegraphics[width=1.2\textwidth]{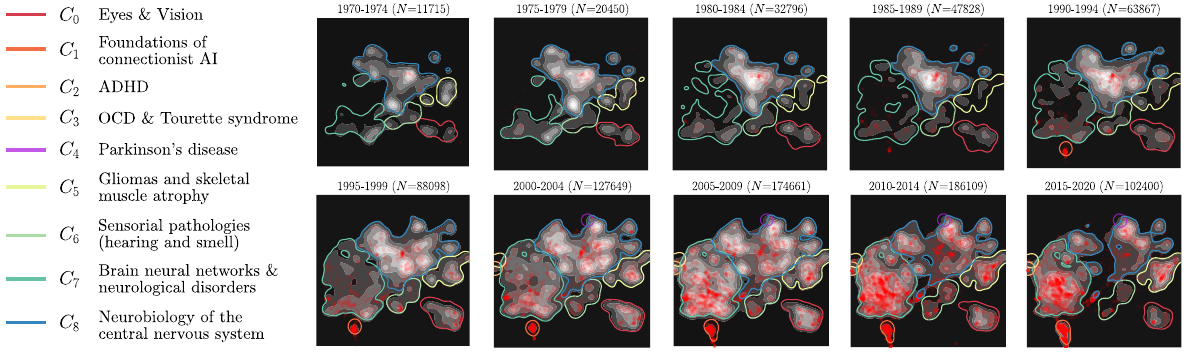}
    \caption{Evolution of the neuroscience knowledge space and its clusters, plotted every 5 years between 1970 and 2020.
    $N$ is the number of publications in the corresponding period.
    The red dots are those related to AI.
    The clusters delineated by colored lines, whose names are given in the legend on the left of this figure, are the same as those delimited in the time-aggregated map in Fig.~\ref{fig:map}.}
    \label{fig:map_temporal}
\end{figure}

We complete the map of Fig.~\ref{fig:map} by plotting its temporal decomposition in Fig.~\ref{fig:map_temporal}, in order to approximate the evolution of the neuroscience subfields and the location of AI in the different configurations shaped by them between 1970 and 2020.
With these maps, we identify three phases in the development of the main field of neuroscience.

The first, spanning the period 1970-1984, marks the consolidation of the rightmost clusters of the time-aggregated map, namely the research around the central nervous system, the skeletal muscle diseases, sensorial pathologies, and eyes and vision -- with a growing number of publications inside them, as already shown in Fig.~\ref{fig:map}.
The second phase, running from 1985 to 2009, shows the lexical expansion of these clusters, the substantial growth of the subfield related to the studies of brain neural networks and neurological disorders, and the birth of the knowledge universe around connectionist AI, especially during the subperiod 1985-1989.
We also observe the unification of the knowledge subspace of all clusters except the studies in vision and foundations of connectionist AI during this second period 1985-2009.

The third and last stage, running from 2010 to 2020, shows a division of the aforementioned continuum of clusters into two parts whose boundaries no longer meet, a first one including the left most clusters and some subsets of the cluster related to the studies of the central nervous system ($C_8$), and a second one including the remaining left-most, fragmented clusters.
This period also signs the decline of the subfields of eyes and vision studies and sensory pathologies.

\begin{figure}[t!]
    \centering
    \includegraphics[width=0.8\textwidth]{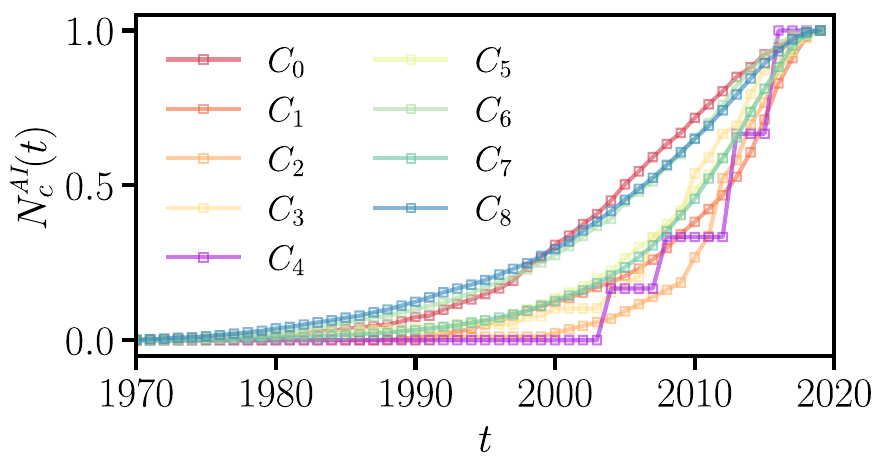}
    \caption{Cumulative number of AI publications per cluster, normalized by their respective maximum value reached in the year 2019 in order to compare their respective growth trend.
    The stepped shape of the $C_4$ curve (Parkinson's disease studies) is due to the very limited number of AI-related publications within it.}
    \label{fig:cumulNb_AIpapers_norm}
\end{figure}

These temporal snapshots of the knowledge space in Fig.~\ref{fig:map_temporal} also suggest an acceleration in the growth of AI research since this period 1985-1989.
This trend is confirmed by the cumulative number of AI publications per cluster shown in Fig.~\ref{fig:cumulNb_AIpapers_norm}.
While studies on vision ($C_0$), sensory pathologies ($C_6$) and central nervous system ($C_8$) exhibit a sustained growth of AI publications within them, all others show a faster growth of such scientific production in them during the 2000s, the most spectacular one having happened in ADHD studies ($C_2$), which has particularly been fast after the 2010s.
The pre-existing knowledge and the dominant vocabulary in these last clusters seem to favor the development of AI within their own research context, for example, by attracting and pairing external AI-related concepts with their own conceptual basis, or by allowing the emergence of other AI-related ones -- which is more difficult to establish.
The next section is dedicated to examine such microscopic aspects related to the organization of knowledge in all clusters of the map.

\subsection{The limited integration of AI into the various concept networks of neuroscience}
\label{sec:intrication}

Through the MAG fields of study assigned to each article in our neuroscience dataset $\mathcal{P}$, which are also identified by their respective knowledge cluster memberships, we delve below into the first dimension evoked in the very last lines of the previous section, namely the entanglement of AI concepts within the conceptual universe of these clusters.

\begin{figure}[t!]
    \centering
    \includegraphics[width=0.9\textwidth]{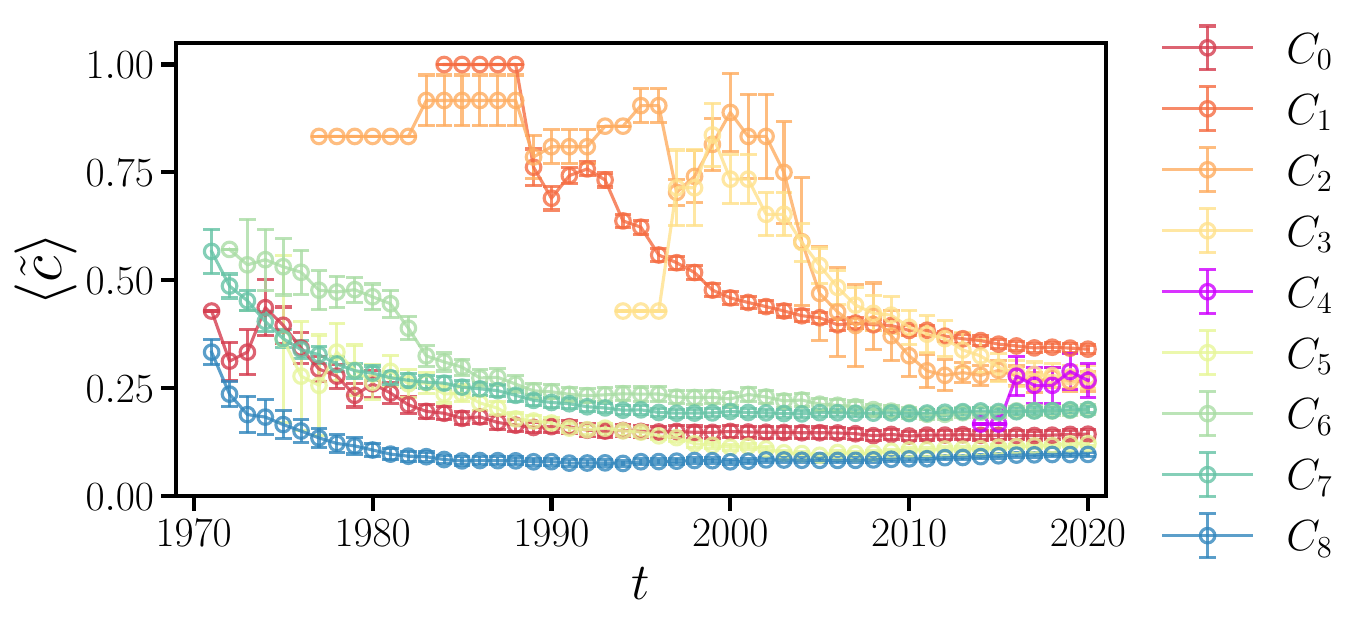}
    \begin{tabular}{|c|c|c|c|c|}
        \hline
        \multirow{2}{*}{\textbf{Cluster}} & \bm{$\langle\tilde{c}\rangle$} \textbf{(year of} & \multirow{2}{*}{\bm{$\langle\tilde{c}\rangle$} \textbf{(2020)}} & \textbf{Number of AI-} & \textbf{Share of AI-}  \\
         & \textbf{apparition)} &  & \textbf{related concepts} & \textbf{related concepts}  \\
       \hline
        $C_0$ & 0.43 (1970) & 0.14 & 209 & 0.02 \\
        $C_1$ & 1 (1983) & 0.34 & 1062 & 0.16 \\
        $C_2$ & 0.83 (1977) & 0.27 & 55 & 0.03 \\
        $C_3$ & 0.43 (1993) & 0.28 & 37 & 0.03 \\
        $C_4$ & 0.17 (2013) & 0.27 & 4 & 0.01 \\
        $C_5$ & 0.37 (1974) & 0.12 & 376 & 0.02 \\
        $C_6$ & 0.57 (1971) & 0.2 & 283 & 0.02 \\
        $C_7$ & 0.57 (1970) & 0.2 & 1245 & 0.04 \\
        $C_8$ & 0.33 (1970) & 0.1 & 590 & 0.02 \\
        \hline
    \end{tabular}
    \caption{Temporal average coreness $\langle\tilde{c}\rangle$ of the AI-related concepts in the cumulative temporal concept network of each cluster -- normalized by the highest coreness returned by the $k$-core decomposition of the network at a given year $t$.
    The higher the coreness $\langle\tilde{c}\rangle$, the closer the AI-related concepts are to the core of the network, and vice versa.
    We only consider AI-related concepts included in the giant component of each cluster's concept network.
    The error bars are the standard errors produced by the distribution of corenesses of all AI-related concepts within each cluster at each year, i.e. the ratio of the standard deviation to the square root of the number of entities present in the distribution.
    In the table below the plots are indicated, for each cluster, the value of the coreness of AI in the cluster's inception year, the final coreness of AI in 2020, the number of unique AI-related concepts in 2020, and the share of such concepts among all the others in 2020.}
    \label{fig:coreness_concepts_AI_cluster}
\end{figure}

Here, we report this articulation by using the cumulative concept network of the clusters in order to track the location of AI-related concepts within it over time -- in the core or in the periphery --, which is summarized by the mean temporal corenesses of AI-related concepts plotted in Fig.~\ref{fig:coreness_concepts_AI_cluster} for each cluster (see Sect.~\ref{sec:build_concept_net} for the computation of this metric). 
More interestingly, it shows a general retreat of AI from the core to the periphery of the concept networks in all clusters over the years, even in the mathematical and computational developments of AI.
In particular, while AI becomes less central over the period 1970-2020 in the respective conceptual framework of the largest knowledge clusters, including the aforementioned papers' continuum and the subfield of eyes and vision studies ($C_0$, $C_5$, $C_6$, $C_7$ and $C_8$), and seems to migrate to a more distant periphery, it suffers a considerable decline in clusters at the left-most periphery of the knowledge map ($C_1$, $C_2$ and $C_3$), where it was rather central at the time of their respective emergence in neuroscience.
This is especially the case for OCD ($C_2$) and ADHD ($C_3$) studies, where AI's coreness decreases sharply over a short period of time since the early 2000s, 20 and 22 years, respectively.
Although AI participated in the creation of these clusters and thus contributed to lay the first milestones of the conceptual universes of these clusters, the progressive addition of new fields of study and connections between them seems to have transformed these conceptual universes and pulled AI away during the last 20 years.
Nevertheless, only AI's coreness is increasing in Parkinson's disease studies ($C_4$) between 2015 and 2016, showing a progressive interest of the latter in AI-related knowledge.

The behavior of the coreness of AI within the core cluster of AI development ($C_1$) is particularly intriguing.
Indeed, while the majority of its publications are based on AI and are limited to a very small vocabulary subspace, as previously shown by the cartography in Figs.~\ref{fig:map} and \ref{fig:map_temporal}, the very numerous AI-related concepts attached to its publications compose only 15\% of the population of unique concepts within this cluster in 2020.
This suggests that this cluster also faces a conceptual reconfiguration, so that the AI concepts that contributed to shape this cluster open up to other, external ones that are not related to AI but close to the knowledge domains centered around the studies of neurological pathologies.
The case of cluster $C_1$, as well as those of $C_2$ and $C_3$ depicted above, thus illustrate the adaptation of AI-centered clusters to larger conceptual universes.

\subsection{The limited citation spreading of AI-related publications in neuroscience}
\label{sec:citation_diffusion}

\subsubsection{AI-related articles diffuse less than non-AI ones across the domain}
\label{sec:citation_diffusion_gyradius}

In this part, we investigate the ability of AI-related articles ($\mathcal{P}\cap AI$) to spread their knowledge across neuroscience by comparing their citation impact in different knowledge subspaces of this field with that of the non-AI articles ($\mathcal{P}\cap \overline{AI}$).

\begin{figure}[t!]
    \centering
    \includegraphics[width=1\textwidth]{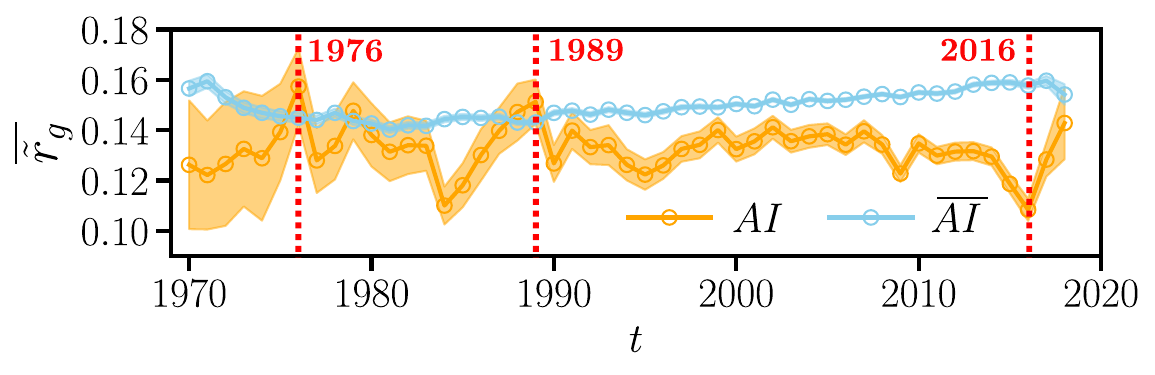}
    \caption{Mean temporal citation radius of gyration of the neuroscience papers, AI-related (orange) and non-AI (blue).
    The colored areas around the curves are their respective standard errors, expressed here as the ratio of the standard deviation of a sample in a given year to the size of that sample.
    We only consider the papers with at least 3 citations.}
    \label{fig:gyradius_citation_average}
\end{figure}

Starting from the citation network between the papers in the dataset, and by selecting only those with at least 3 citations, we compute for each of them the spatial radius of gyration (RoG) produced by the positions of their respective citing papers in the knowledge space, following the method introduced in Sect.~\ref{sec:gyradius_calculation}.
We average these measures per year, distinguishing the AI-related papers from the non-AI ones, as shown in Fig.~\ref{fig:gyradius_citation_average}. 

The average RoG of the citations received by the non-AI papers, which first decreases slightly until the 1980s, increases almost linearly since this period, while the evolution of the measure applied to the AI-related papers fluctuates with high amplitudes\footnote{This behavior could be explained by the very heterogeneous distribution of these papers during the studied temporal period -- 87\% of the papers are published after 1990 -- and their high dispersion in the conceptual space -- as shown by a standard error that is significantly higher before the 1990s than after.}, especially before the early 1990s.
Overall, the latter remains rather below the former after the 1990s, thus meaning that AI-related articles diffuse less than the non-AI ones within neuroscience over the time period studied.

\begin{figure}[t!]
    \centering
    \includegraphics[width=1\textwidth]{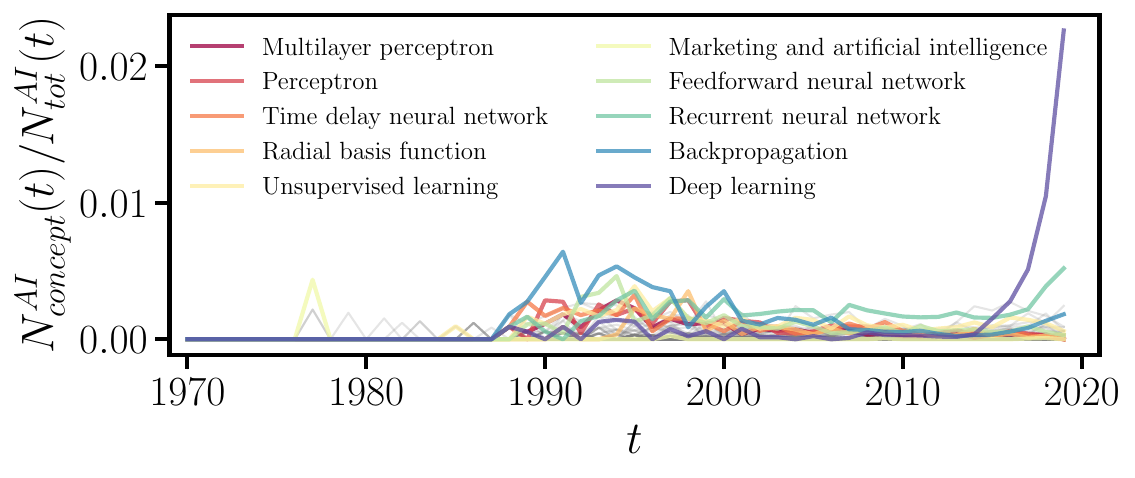}
    \caption{Share of AI-related fields of study provided by MAG among the AI-related articles in neuroscience over the years. 
    The top 10 are highlighted in colors, others are in gray.
    The concept \textit{Marketing and artificial intelligence} is inherited from another one called \textit{Intelligent Decision Support System}, namely a subfield of symbolic AI that was \textit{in vogue} between the 1970s and the 1990s.
    The algorithms populating this class of AI are based on complex knowledge representations in order to provide some assistance to decision makers. 
    One of the most important algorithms in the medical and clinical domains during the aforementioned period, including neuroscience, was the so-called \textit{Mycin}, which is also widely mentioned in the AI-related papers of our database during the late 1970s.}
    \label{fig:AIkw_evol}
\end{figure}

Following the average citation RoG of the AI-related articles, the lexical coverage of the publications citing the AI-related work between the 1970s and the 1990s alternates between phases of low (pits) and high (peaks) citation diffusion.
In particular, the curve shows two peaks higher than the curve of the non-AI papers in 1976 and 1989.
These peaks correspond to two well-known stages in the history of AI, as shown by the evolution of the use of AI-related concepts in Fig.~\ref{fig:AIkw_evol}, namely the symbolic and the connectionist ones.
In particular, we notice the rise of \textit{Backpropagation} and stochastic gradient descent techniques on artificial neural networks in the late 1980s, especially popularized by Rumelhart et al. (\citeyear{rumelhart_learning_1986}), and that is directly inherited from the reinvestment of the concept of \textit{Perceptron} -- single or multilayer \citep{rosenblatt_perceptron_1958}.
This will lead to the rise of neural network techniques and later to deep learning ones \citep{cardon_neurons_2018}.
This connectionist phase fosters the development of neuroscience through the formalization of cognitive processes under (artificial) neural networks, which are useful to study the development of neurological disorders in recent days (see Section~\ref{sec:map}).
Some further results, detailed in App.~\ref{app:dynamic_gyration_radius}, also show that the related findings during the two studied periods have achieved a kind of success, as testified by the rapid citation spread of AI-related ideas across the neuroscience knowledge space.

However, we notice in Fig.~\ref{fig:gyradius_citation_average} a strong decrease in the RoG between 2014 and 2016, meaning that the AI-related knowledge and tools embedded in these papers are beneficial only to their close lexical neighborhood.
In fact, among the 3,739 papers published between 2015 and 2016 during the decrease in the mean RoG, 37\% emanates from studies on neurological disorders and brain neural networks ($C_7$) and 35\% from the subfield developing connectionist AI knowledge and tools ($C_1$), while the rest is distributed among the other clusters and noise from the knowledge map.
This distribution supports one of the phenomena already evoked by Fontaine et al. (\citeyear{fontaine_epistemic_2024}), namely the specialization of AI research around the subfield designing AI itself in neuroscience (cluster $C_1$), accompanied by other applicative subfields mainly confined to neurological disorders and brain neural networks studies (cluster $C_7$).

The late decrease in the mean RoG associated with AI-related concepts is particularly intriguing because it falls within a period of well-documented, high diffusion of AI in science, associated with the rise of deep learning techniques \citep{cardon_neurons_2018,gargiulo_meso-scale_2023}, which reasonably suggests that any paper using or mentioning such concepts would find a wide audience throughout the whole knowledge landscape of neuroscience.
We will see in the next subsection that this decrease and, more generally, the low RoG of AI-related articles are parts of a larger phenomenon of concentration of AI-related neuroscientific knowledge within the clusters from which they originate, which does not guarantee the epistemic genericity of AI within the domain under study.

\subsubsection{AI remains confined to local knowledge subspaces of neuroscience}

\begin{figure}[t!]
    \centering
    \includegraphics[width=\textwidth]{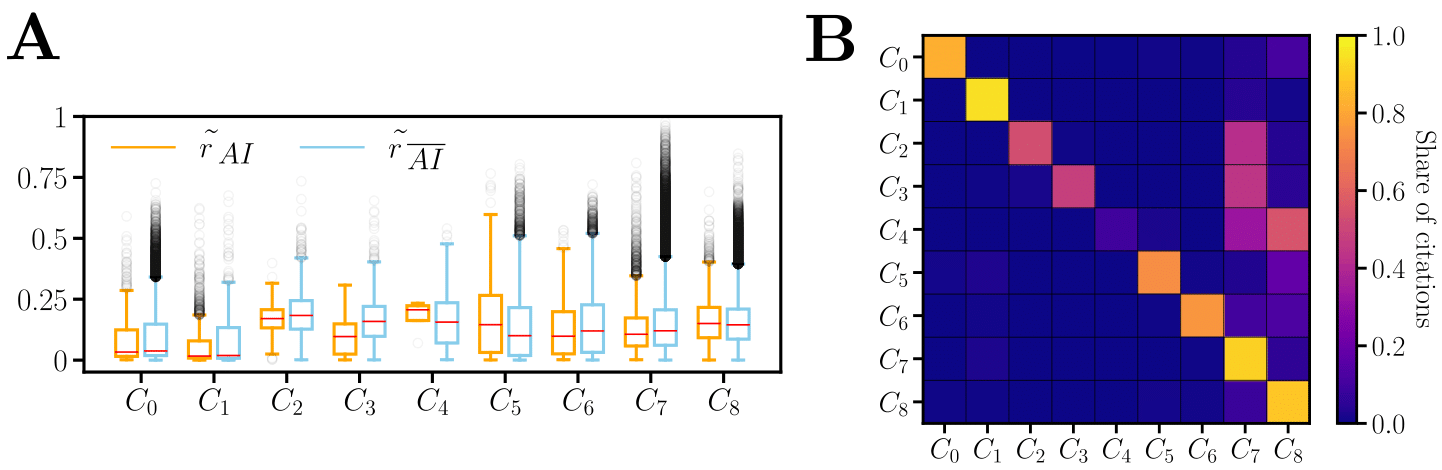}
    \caption{\textbf{A}: Time-aggregated distribution of the citation RoGs of the neuroscience papers within each cluster, AI-related (orange) and non-AI (blue).
    The red lines in the boxes indicate the medians of the distributions.
    Here, we only consider the papers with at least 3 citations.
    \textbf{B}: AI-centered citation matrix between the clusters of the neuroscience knowledge map shown in Fig.~\ref{fig:map} (between 1970 and 2019).
    A cell ($C_i,C_j$) is the number of papers in $C_i$ that cite the AI-related articles in $C_j$.
    The results are normalized by row, so the sum of the score for a row is 1.}
    \label{fig:gyradius_citation_distrib_cluster}
\end{figure}

As suggested in the previous subsection, the spreading of the AI instrument across the entire neuroscience knowledge space remains quite limited over the years. 
In this section, we explore at the microscale the clustered distribution of the citation RoGs of all articles in our dataset, as well as the time-aggregated citation network centered on the AI-related works and aggregated into clusters.
Both are represented in Fig.~\ref{fig:gyradius_citation_distrib_cluster}.

According to Fig.~\ref{fig:gyradius_citation_distrib_cluster}A, which complements the previous Fig.~\ref{fig:gyradius_citation_average}, the publications dealing with AI within all the clusters of the neuroscience knowledge map diffuse as much or less than the non-AI ones, except for the small peripheral clusters on studies of Parkinson's ($C_4$) and skeletal muscle diseases ($C_5$), which both show the highest citation RoGs for the AI knowledge produced in them.
This low diffusion is especially observed in the clusters where AI has been mainly developed in recent years (see Section~\ref{sec:map}), namely studies on connectionist AI development ($C_1$), ADHD ($C_2$), OCD and Tourette syndrome ($C_3$) and other neurological disorders ($C_7$).

The AI-centered citation matrix in Fig.~\ref{fig:gyradius_citation_distrib_cluster}B also shows a strong phenomenon of self-citation within each cluster, except for the small peripheral clusters $C_2$, $C_3$, and $C_4$, which are also influenced by the largest clusters $C_7$ and $C_8$ that constitute the core of the knowledge map.
The AI-related knowledge produced in a given cluster thus preferentially impacts that cluster -- and/or its neighbors -- and not the others.

\section{Discussion}
\label{sec:4_discussion}

How does AI-related knowledge develop and spread in neuroscience?
Through the analysis of a dynamic knowledge map of neuroscience generated with advanced document embedding techniques applied to the textual elements (titles and abstracts) within our corpus, which are also connected by citations, we have shown that AI is found in every lexical regions of the map, but it meets a greater success within subfields that have begun to grow significantly since the 1990s, and that are related to the mathematical and computational foundations of contemporary connectionist AI, and to the studies of the human brain neural network and various neurological disorders that lead to physiological and/or cognitive impairments.
In particular, the presence of the majority (59\%) of the total number of AI-related publications in these regions tends to confirm the implantation of artificial neural network architecture as an underlying paradigm in the most recent research domains of neuroscience.
These first findings provide some empirical evidence to support the considerations of the neuroscientists themselves about the potentialities of applying AI to their own research goals, and especially about the reinforcing feedback between the host research field and the AI instrument towards their mutual development \citep{macpherson_natural_2021,gopinath_artificial_2023,hassabis_neuroscience-inspired_2017,nmi_neuroAI_2024}.

In an analysis of the set of fields of study labeling the papers within each knowledge cluster of the map, which constitute the delimited conceptual framework of the latter, we have found that AI is not integrated in the various conceptual arcs of the domain, except for the small, dense, constantly isolated knowledge subspace related to the formal conception of connectionist AI and computational neuroscience, which includes 22.5\% of the AI-related production in our corpus -- and which also exhibits a strongly confined citation network, as shown in Fig.~\ref{fig:gyradius_citation_distrib_cluster}B, similar to an echo chamber.
Finally, this cartography supports the argument of the dual epistemic orientation of AI research in the discipline that has been previously shown by Fontaine et al. (\citeyear{fontaine_epistemic_2024}): The aforementioned island of knowledge would represent the epistemic environment in which AI serves the purposes of the so-called STEM disciplines, while the rest of the AI-related publications spread across the entire map represent its epistemic orientation towards neuroscience itself, thus demonstrating a certain epistemic genericity of AI throughout the entire domain \citep{shinn_transverse_2002,marcovich_regimes_2012}.

However, although AI is one of the cornerstones of the conceptual edification of all the neuroscience subfields, it progressively moves away from the core of all the concept networks representing these subfields as they expand over time.
This result thus challenges the seemingly widespread distribution of AI exhibited by the knowledge mapping.
Indeed, the lexical proximity of AI-related articles to other works within each cluster, confronted with the organization of the concepts attached to them, does not allow us to firmly establish the genericity of AI produced in neuroscience.
This finding leads instead to support the fact that AI is a general method that could be applied everywhere in neuroscience \citep{cockburn_impact_2018}, without being a source of new concepts that could enrich the pre-existing conceptual frameworks of all its subfields.

Furthermore, with the citation network between the articles, we have constructed a measure of the citation scope of an individual paper, the citation radius of gyration (RoG), which enables us to dynamically identify some time periods when its encoded knowledge has or has not spread beyond its local lexical neighborhood.
In particular, we have unveiled alternating phases of increase and decrease in the average citation scope depicted over time that coincide with well-known events that structure the evolution of AI and its impact on science, such as the development of expert systems in the late 1970s and the rise of artificial neural networks in the late 1980s.
Although AI is present in all of the neuroscience research subfields revealed by our data, we have shown that these subfields locally disseminate their own AI publications, with very little circulation of knowledge between the subfields. 
Thus, AI is only beneficial for the neuroscience subfields in which it has been produced.
This is especially the case for machine learning or deep learning models and algorithms, which are highly dependent on the data used to train them before performing predictions for very precise purposes.
These assumptions also indicate that the applications of AI remain localized in specific topics and do not seem to be transferable to other knowledge subspaces of neuroscience, thus suggesting the multiplication of distinct AI research dedicated to different purposes.
For example, a complex AI architecture shaped and trained to predict the course of Alzheimer's disease in a patient of a certain age would also not be able to predict the progression of Parkinson's or Huntington's disease.


In summary, the vast distribution of scientific articles involving AI on the drawn map indicates an apparent epistemic genericity of the related knowledge and technologies across neuroscience, while being less and less aligned with the conceptual universes the of neuroscience subfields as they develop, as we have shown with conceptual structures based on the co-occurrence network of fields of study provided by MAG and that label the papers of our dataset.
AI thus demonstrates its ability to be applied in all the subfields delimited in our study, which include various knowledge areas for the largest one, but it does not intend to create new knowledge and conceptual universes within them, except for a small lexical region that is dedicated to it.

These contrasting findings complete the perspectives evoked in a previous paper \citep{fontaine_epistemic_2024}, according to which AI is well inserted in the multidisciplinary context of neuroscience, while its practitioners, regardless of their level of expertise, experience together a segregation within the whole field.
Here, AI seems to be generic \textit{in application}, as demonstrated by its spread in various lexical areas covering the knowledge space of neuroscience, which is a signal of its adaptation in almost all epistemic frameworks of the domain, but not generic \textit{in conceptualization}, i.e., its \textit{metrology} is limitedly disseminated, as it withdraws from their conceptual arcs that could give rise to new theories and research paradigms.
Paradoxically, AI seems to lose this second genericity over time in neuroscience and consequently does not seem to replace all the methodological frameworks present in the field.

\subsection*{Limitations and future work}

A major limitation of this study lies on the clustering provided by HDBSCAN (see Fig.~\ref{fig:map}) and its effect on the citation RoG.
Although it remains relevant for studying the confinement of AI-oriented citations within large lexical areas, as shown in Fig.~\ref{fig:gyradius_citation_distrib_cluster}B, the RoG distributions in Fig.~\ref{fig:gyradius_citation_distrib_cluster}A, on the contrary, suggest that very different lexical subareas, with their own and probably disjoint citation networks (composed of papers with very low RoG), also coexist within these clusters, like small specialties \citep{wray_rethinking_2005}.
It would then be worth refining the clustering in order to better observe the bridges between smaller lexical areas represented by few articles, such as the small clusters on ADHD, OCD and Parkinson's disease studies in Fig.~\ref{fig:map}, whose AI-related publications find interest in nearby lexical areas within the larger clusters on brain neural networks, various neurological disorders and central nervous system studies.
Moreover, we have focused on the direct citations of AI-related publications, but we could imagine a more complex dynamics of impact involving citation chains towards the different clusters in time, from which we could compute the number of citations needed to reach a particular cluster from another, and evaluate potential bifurcations through fields outside neuroscience before reintegrating the latter -- following the exemplary frameworks depicted in \citep{brahim_data-driven_2021,huang_number_2018}.

Furthermore, the progressive disembedding of AI from the core of all conceptual frameworks of neuroscience (see Fig.~\ref{fig:coreness_concepts_AI_cluster}) suggests that a social component involved in this AI-related research is progressively pushed to the periphery of the collaboration network, thus testifying the rise of a high specialization of AI towards a more formal and technology-oriented epistemic orientation, as also depicted in \citep{klinger_narrowing_2022} for the global science system.
Although we have not addressed here the location of the neuroscientists on this map, a further work would consist of projecting the co-authorship network onto this map -- by locating the authors with an aggregated position derived from their respective publications -- such that we could identify the knowledge subspaces in which the AI practitioners are located, and clearly validate or not the aforementioned hypothesis of a specialization of AI practitioners towards a dedicated conceptual framework distinct from the rest of neuroscience.
Such a mapping would also allow us to delineate specific socio-epistemic environments in neuroscience, i.e., sub-networks of collaborations organized around a common knowledge base, so that we could trace the potential social bridges between them that would promote the diffusion of particular knowledge from one epistemic community to another, in particular AI, and also identify the actors involved in such bridges.\footnote{We refer to the work of Herfeld and Doehne (\citeyear{herfeld_role-typo_2019}), which proposes a classification of the different roles endorsed by scientific contributions within a diffusion process on a citation network. This classification could be relevant in describing the degree of specialization in AI of neuroscientists themselves, and how it evolves over time.}
As in \citep{roth_social_2010}, we could also build a socio-conceptual network linking neuroscientists to the fields of study related to their articles, in order to delineate the conceptual universe that is preferred by the AI practitioners, given their level of expertise on the notion and their membership cluster.


\section{Declarations}

\begin{itemize}
    \item Ethics approval and consent to participate: Not applicable.
    \item Consent for publication: Not applicable.
    \item The datasets used for the this study are available at: \url{https://doi.org/10.5281/zenodo.17589028}. The codes for analyzing them are available at: \url{https://github.com/sysyMC/AI_in_Neuroscience_TopicModeling}.
    \item Competing interests: The authors declare that they have no competing interests.
    \item Funding: S.F. is funded through a CNRS-MITI PhD grant within the project ``Epistemic Impact of Artificial Intelligence in Science'' (EpiAI). 
        This research has been partially supported by the ANR grant ScientIA (ANR-21-CE38-0020).
    \item Authors' contributions: S.F. conceived the research, collected and analyzed the data, discussed the results, wrote the manuscript.
    \item Acknowledgements: S.F. warmly thank M. Dubois, F. Gargiulo and P. Tubaro for useful discussions.
\end{itemize}

\begin{appendices}

\section{Technical elements for the construction of the neuroscience knowledge map}
\label{app:knowledgeMap_construction}

\subsection{Extraction of supplementary metadata from OpenAlex}
\label{app:OA_extraction}

For the special needs of interpreting the neuroscientific knowledge map whose building and interpretation is detailed within the main text, we complete the MAG's conceptual classification with another one launched in early 2024 by OpenAlex, according to which each paper in the database is labeled with a \textit{primary topic} and two others (sorted by a matching score). 
These topics are leaves of a four-level hierarchical classification that is actually a tree:\footnote{See also: \url{https://docs.openalex.org/api-entities/topics}.} Level 0 is composed of domains, each of which is divided into fields at level 1, which are subdivided into subfields at level 2, which contain a given set of topics at level 3. 
More synthetic than the MAG's fields of study, this topical taxonomy will be used in the next sections to confirm some of the results initially obtained with the former, especially the thematic consistencies of the SPECTER embedding and the UMAP one analyzed above.
In particular, on 30 June 2024, we have performed an extraction from OpenAlex of such topics associated with the papers in our corpus, and we conserve only the primary one for each of them -- with their own respective subfield, field, and domain.
We also extracted from OpenAlex the \textit{keywords} describing the papers.

\subsection{Robustness of the UMAP projection with $k$NN procedures applied with OpenAlex metadata}
\label{app:knn}

With the SPECTER model \citep{cohan_specter_2020}, we have converted the titles and abstracts of the articles in our neuroscience corpus into 768-dimensional vectors summarizing their respective lexical features and scientific knowledge.
Then, we have projected these high-dimensional vectors into a latent two-dimensional space using the UMAP method \citep{mcinnes_umap_2020}.
As mentioned in Sect.~\ref{sec:build_map}, the projection made by UMAP is not unique.
In this appendix, we test the robustness of this projection with two kinds of \textit{nearest neighbors research}, both of which allow to verify whether the points preserve their respective neighbors from the high-dimensional space to the low-dimensional one under study in this manuscript.

The first one merely consists in running such a nearest-neighbors search for each paper of the dataset and comparing its two respective ensembles of neighbors in the two embedding spaces -- hereafter simply called SPECTER and UMAP.
More precisely, we perform this comparison by computing for each paper its share of common neighbors exhibited in these two spaces.
With the module \texttt{NearestNeighbors} from the Python package \texttt{scikit-learn}, with a neighborhood population to query set to $k=100$ (which is not too large to avoid covering entire regions of the map) and a calculation mode set to pairwise Euclidean distances, we find an average conservation of 10.8\%.
This very huge loss after projection, however, does not imply that the papers are not surrounded by other neighbors in the same area of vocabulary and encoded knowledge (or nearby).
That is why, in a second time, we verify whether the articles remain in a similar thematic environment after the UMAP projection.

Inspired by the approach given by Singh et al. (\citeyear{singh_charting_2024}) and González-Márquez et al. (\citeyear{gonzalez-marquez_landscape_2024}) in their respective technical materials used to scientometrically reconstruct broad research landscapes, we implement a $k$-nearest neighbors ($k$NN) procedure to decipher the thematic structure of the SPECTER-embedded corpus and the UMAP-embedded one.
For a given set of points, such a $k$NN classifier predicts their unique class given the classes labeling their respective neighbors, i.e. it assigns the majority class appearing in the neighborhood of the point under study.
To achieve this, we choose to label the papers with their respective \textit{subfield} associated with their \textit{primary topic} extracted from OpenAlex, specifically for two reasons: 1) the \textit{topics} are too numerous in the dataset (2881), and 2) the disciplinary or thematic classifications used throughout the main text of the manuscript label the articles with neither a unique discipline (from WOS JSCs) nor a unique field of study (from MAG), thus compromising the execution of a $k$NN search.
With such a subfield classification, the application of $k$NN to the SPECTER and UMAP datasets allows to build two confusion matrices summarizing the proportion of true predictions per subfield, which will be compared afterwards.

\begin{figure}[t!]
    \centering
    \includegraphics[width=0.8\textwidth]{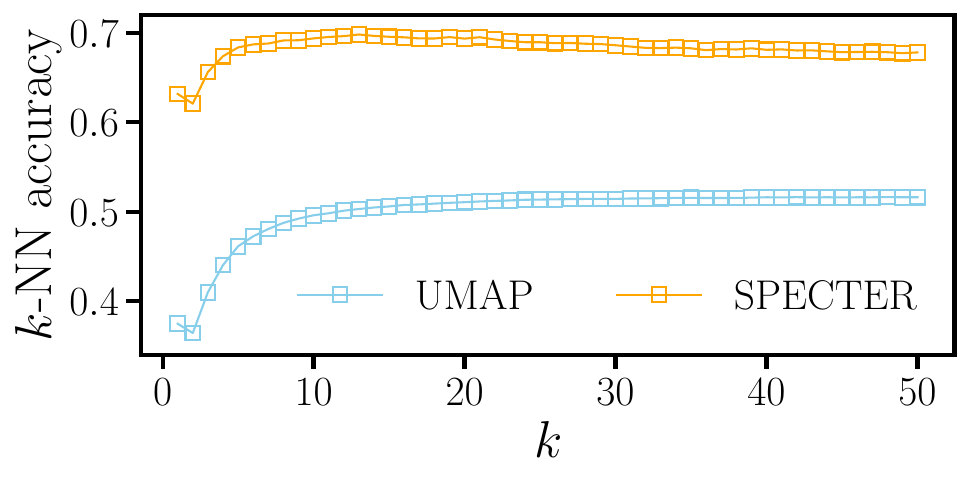}
    \caption{Accuracy of subfield prediction given by a $k$-NN algorithm at several number of neighbors $k$ for each embedding space -- 768 dimensions for SPECTER and two dimensions for UMAP. 
The accuracy at a given $k$ is computed as the share of points in the test set whose labeling subfield prediction is correct, i.e. the predicted subfield matches the actually observed one.
For SPECTER, 99\% of the dataset was used to train the algorithm at each $k$, and the remaining 1\% served to compute the accuracy.
For UMAP, we choose 80\% of the dataset to train the algorithm, and 20\% to test it.}
    \label{fig:accuracy_knn}
\end{figure}

In what follows, we restrict our analysis to a subset of papers covering 68 subfields, each labeling at least 1000 publications -- which accounts for 96\% of the original dataset.
We thus keep their associated vectors within the two embedded datasets.
Before plotting their respective confusion matrices, we choose for each of them the number of neighbors $k^*$ maximizing the accuracy of the predictions returned by the search algorithm.
As shown in Fig.~\ref{fig:accuracy_knn}, the accuracy for UMAP is limited to about 51\% from approximately $k^*=20$, while SPECTER's one reaches its maximum of about 70\% at $k^*=12$.
We choose these values to run optimized $k$NN searches within each dataset -- with the help of the module \texttt{KNeighborsClassifier} of the Python package \texttt{scikit-learn}.

\begin{figure}[t!]
    \centering
    \includegraphics[width=\textwidth]{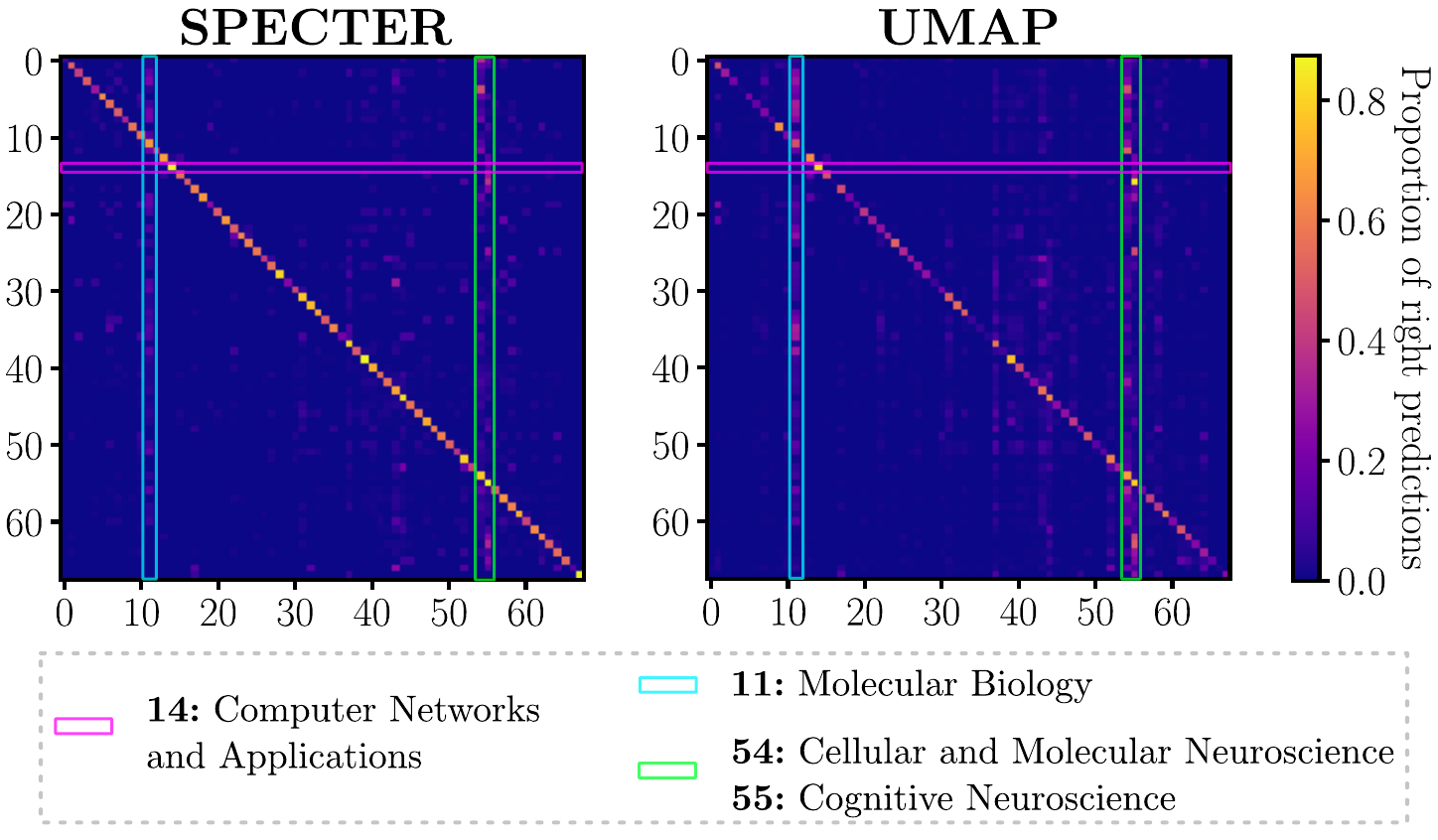}
    \caption{Confusion matrix of $k$NN predictions per subfield for the SPECTER-embedded dataset (left) and the UMAP-embedded one (right).
    The indices of the matrices represent the 68 subfields considered to run the $k$NN models.
    The rows are the actual observed subfields and the columns are the predicted ones, sorted as in the rows.
    Given a subset of papers labeled with an observed subfield $i$ (indexed in rows) within the test set, the cell $m_{ij}$ in each matrix is the share of predictions falling within the subfield $j$ (indexed in columns).
    For example, among the test papers labeled with the observed subfield \textit{Computer Networks and Communications} (pink-colored row 14 in the heatmaps) in the SPECTER space, 87.5\% of them have a correctly predicted subfield, i.e. the observed one, but the remaining 12.5\% have a predicted subfield other than their common observed one, here \textit{Cognitive Neuroscience}. 
    In the UMAP space, this is 81\% of them that are correctly predicted into the observed aforementioned subfield, against 19\% exhibiting other subfields -- 14 in total. 
    Therefore, the sum of the prediction shares equals 1 for each observed subfield indexed in rows.
    For SPECTER, 99\% of the dataset was used to train the $k$NN algorithm with $k^*=12$, and the remaining 1\% served to compute the predictions (with an accuracy of about 70\%).
   For UMAP, we choose 80\% of the dataset to train it with $k^*=20$, and 20\% to compute the predictions (with an accuracy of about 51\%).
   The light blue rectangle corresponds to the columns associated with the subfields predicted as \textit{Molecular Biology}, and the light green one to the subfields predicted as \textit{Cellular and Molecular Neuroscience} and \textit{Cognitive Neuroscience}.}
    \label{fig:confusion_matrices}
\end{figure}

The results of the predictions per subfield are shown in Fig.~\ref{fig:confusion_matrices}.
They indicate that the predictions given by the $k$NN run over the papers embedded with SPECTER are concentrated around the diagonal of the corresponding confusion matrix, thus testifying that the papers lying in a given subfield are statistically surrounded by neighbors also labeled with that subfield. 
The papers labeled with a common subfield are thus statistically included in a common area of knowledge represented by a common vocabulary, which confirms the rather good embedding performance of SPECTER.
We recover this trend in the projected UMAP space, but with an accentuation of incorrect predictions in some zones of the confusion matrix, corresponding to those already observed in the SPECTER one, but with a lower intensity.
This lack of precision concerns almost all the subfields, which show an important part of the predictions towards the subfields \textit{Molecular Biology} (light blue rectangular zone on the two heatmaps), \textit{Cellular and Molecular Neuroscience} and \textit{Cognitive Neuroscience} (grouped in a light green zone on the heatmaps).
The neighborhoods of articles belonging to the subfields with high prediction rates to the three aforementioned ones are therefore predominantly populated by at least one of these three subfields, especially in the UMAP space.
This suggests, first, that these major subfields cover a large vocabulary basis and therefore a broad part of the neuroscience knowledge landscape, and, second, that they are lexically close to articles labeled with subfields showing them as main predictions in the confusion matrices, such as \textit{Biochemistry} (row 4), \textit{Physiology} (row 12), \textit{Developmental and Educational Psychology} (row 62) and \textit{Experimental and Cognitive Psychology} (row 63).
Moreover, because they are the most populated subfields -- 15\% labeled as \textit{Cognitive Neuroscience}, 14\% as \textit{Cellular and Molecular Neuroscience}, and 10\% as \textit{Molecular Biology}, with the remaining 71\% almost equally distributed among the 65 others -- they are more likely to be predicted than any of the other subfields.

In summary, with a $k$NN algorithm applied to the prediction of the subfields labeling the articles in our neuroscience dataset, we confirm the robustness of the SPECTER embedding of the textual metadata of these articles, and also of the UMAP dimension reduction algorithm used to transform the 768-dimensional vectors representing them into two-dimensional ones.
Albeit a loss of prediction accuracy before and after the projection, we have demonstrated a global conservation of the thematic structure of our corpus, thus enabling us to use the reduced vectors in order to study the cartography of neuroscience shown in Fig.~\ref{fig:map}, and also the location of AI-related knowledge in this reduced space.

\subsection{HDBSCAN dendrogram}
\label{app:dendogram}

In this section, we detail the method retained to partition the knowledge map shown in Fig.~\ref{fig:map} into 9 clusters.

\begin{figure}[t!]
    \centering
    \includegraphics[width=\textwidth]{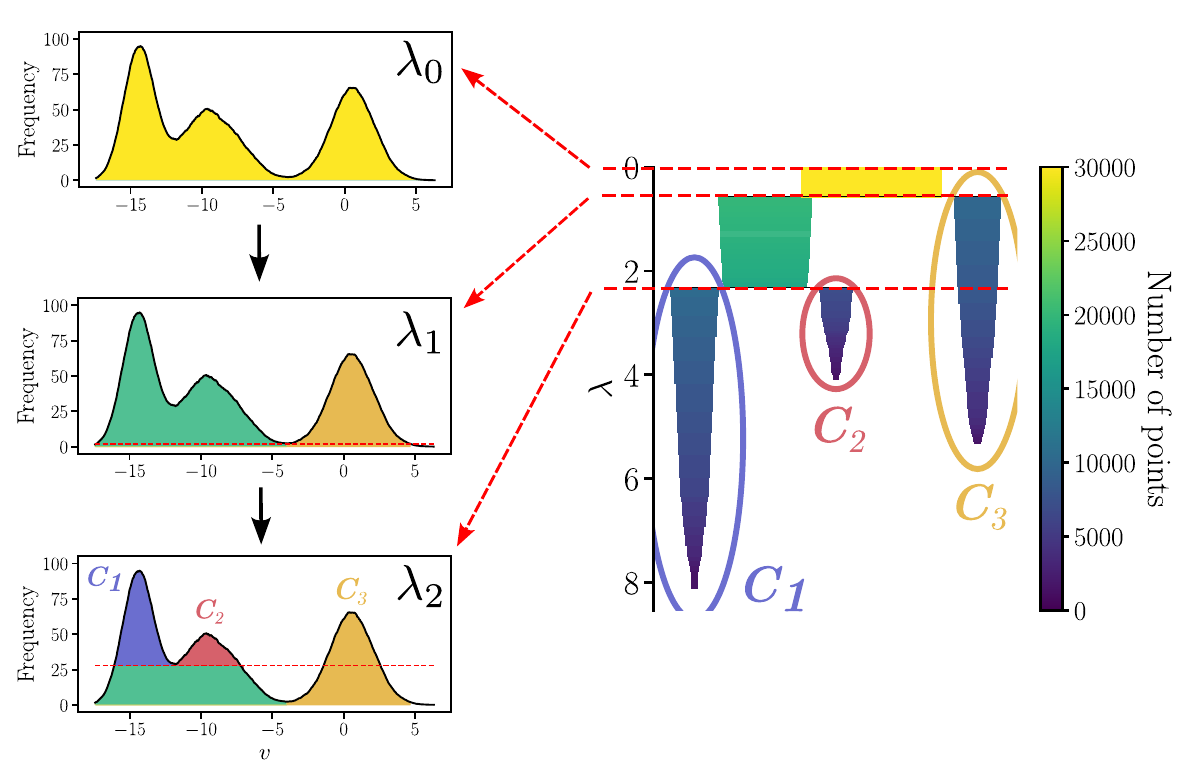}
    \caption{Schematic illustration of the application of HDBSCAN to a set of 30,000 continuous numbers $v$ on a horizontal axis generated from a non-trivial distribution, represented on the left plots.
    This distribution is composed of density peaks corresponding to high amounts of points in a small interval of the horizontal axis $v$, e.g. the leftmost peak (cluster $C_1$) corresponds to a high frequency of values $v$ picked around $-15$.
    On the contrary, the pits are regions with less selected values $v$.
    On the right is plotted the clustering dendrogram returned by the Python module \texttt{hdbscan}, where the density threshold parameter is denoted by $\lambda$. 
    The clustering has been executed with a minimum neighborhood size of 1000 values for each number and a minimal cluster size of 1000 numbers.
    A branch is a cluster whose color is the number of values it contains at a given $\lambda$ -- represented also by the varying width when moving down the tree -- and whose length is the persistence of the cluster as $\lambda$ increases.
    For example, the cluster $C_2$ is less persistent than the other two clusters $C_1$ and $C_3$, i.e. the former is merged with the sea of noisy points at $\lambda\sim 4$, while the latter continue to exist after that value, but with smaller respective sizes.
    As $\lambda$ still increases, $C_3$ is also considered as noise by the algorithm before $C_1$.
    At a value of $\lambda=\lambda_0=0$ of the density parameter, we consider all the points of the distribution, which together shape one unique cluster, colored yellow in the upper left plot. 
    At a value of $\lambda=\lambda_1$, the set of points is split into two clusters, namely the cluster $C_3$ and another one colored in green in the middle plot.
    The latter is finally divided into two subparts at $\lambda=\lambda_2$, namely the clusters $C_1$ (in purple in the bottom plot) and $C_2$ (in wine red).}
    \label{fig:dendrogram_scheme}
\end{figure}

We have drawn upon the main method provided by the Python package \texttt{hdbscan} \citep{mcinnes_hdbscan_2017}, which returns a dendrogram indicating the different clusters we could get by varying a density parameter, denoted here as $\lambda$.
Fig.~\ref{fig:dendrogram_scheme} illustrates a simple use case of this method, here applied to a set of numerical values picked from a heterogeneous distribution.
According to such a dendrogram, choosing a high $\lambda$ is equivalent to selecting small, local and dense ensembles of points, such as those located at the very ends of the branches of the clusters $C_1$ and $C_3$, which correspond to the purple and ochre peaks, respectively, on the lower left plot of the distribution for $\lambda=\lambda_2$.
On the contrary, choosing a small $\lambda$ is equivalent to selecting the largest clusters obtained by merging all the smallest clusters at the bottom of this hierarchy, such as the green cluster in the middle left plot at $\lambda=\lambda_1$, which is the sum of the clusters $C_1$ and $C_2$, and also the yellow one at $\lambda=\lambda_0$ in the top left plot, which is the sum of all the three clusters.
In such a case, the algorithm covers a large amount of points and thus reduces the amount of noise.

\begin{figure}[t!]
    \centering
    \includegraphics[width=\textwidth]{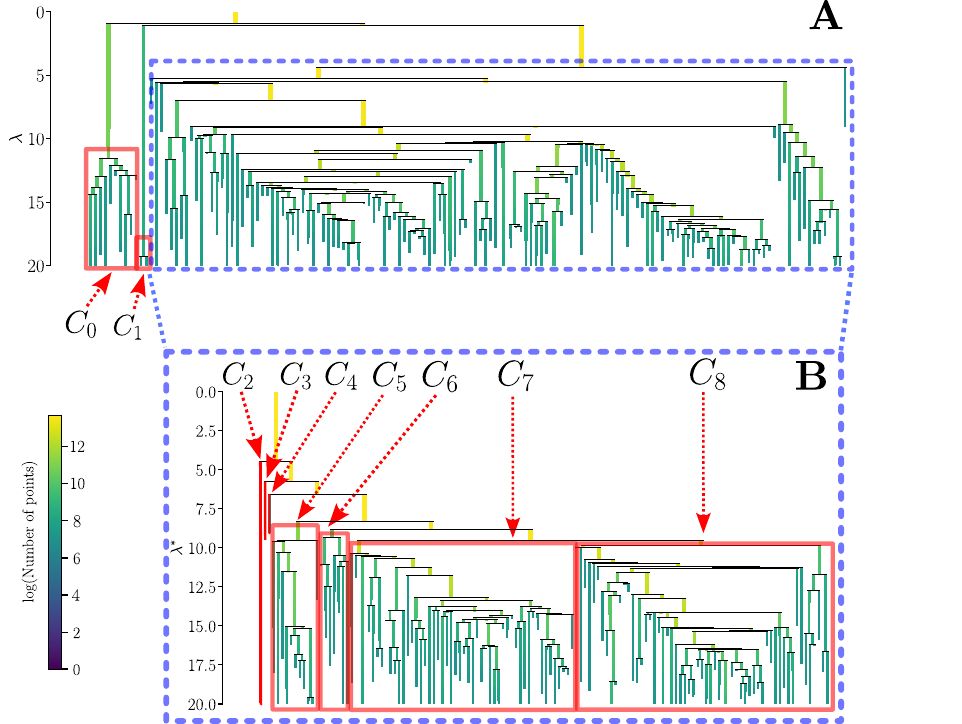}
    \caption{A: Clustering dendrogram returned by HDBSCAN on our reduced dataset. 
    The program has been executed with a minimum neighborhood size of 100 points for each paper and a minimal size of a cluster of 1000 papers.
    The density threshold is denoted as $\lambda$.
    A vertical bar is a cluster, whose color represents the number of points within the corresponding cluster, scaled with the logarithmic color bar on the left side, and whose length represents the persistence of the corresponding cluster according to $\lambda$, i.e. the height of the associated amount of points in the density map.
    B: Clustering dendrogram of the large cluster from dendrogram A above, replotted after a second run of HDBSCAN on the subset of data included in this cluster.
    The density threshold is denoted here as $\lambda^*$ because it is a translated value of the $\lambda$ parameter from plot A -- the lowest value $\lambda^*=0$ actually corresponds to $\lambda\sim 1$. 
    The clusters $C_2$, $C_3$ and $C_4$ are simple bars, colored red for clarity.
    Population of each cluster: $N_{C_0}=56,644$; $N_{C_1}=8,755$; $N_{C_2}=2,222$; $N_{C_2}=1,407$; $N_{C_4}=1,179$; $N_{C_5}=66,465$; $N_{C_6}=28,382$; $N_{C_7}=266,593$; $N_{C_8}=370,814$.
    We also count 55,230 papers that are unclassified by HDBSCAN, i.e. that are considered as noise by the clustering algorithm.
    }
    \label{fig:dendrogram}
\end{figure}

We apply this clustering method to our embedded 2D representation of neuroscience papers, i.e. after applying UMAP.
Fig.~\ref{fig:dendrogram}A shows the dendrogram returned by its execution.
We manually choose the most separated and persistent clusters on this dendrogram, regardless of their respective size, shape, or intrinsic density, by progressively increasing $\lambda$.
When setting $\lambda$ around 5, we divide the dataset into 3 clusters, namely $C_0$, $C_1$ and a very large one whose partition is shown below in Fig.~\ref{fig:dendrogram}B.
In this way we obtain the other clusters appearing in Fig.~\ref{fig:map}.
The populations of all the clusters are indicated in the caption of Fig.~\ref{fig:dendrogram}.
With such a partition, we allow a total noise rate of about 6.5\%.
We considered these time-aggregated clusters for the analysis of all the neuroscience knowledge maps shown throughout the main text, in particular the 5-year temporal maps in Fig.~\ref{fig:map_temporal}.

Furthermore, choosing a high number of neighbors to perform HDBSCAN leads to very aggregated clusters with inhomogeneous density within them, while choosing a low number of neighbors favors the formation of smaller and more homogeneously distributed clusters of points -- hence the very high number of branches Fig.~\ref{fig:dendrogram}.
We have opted for this second strategy to plot this dendrogram for two reasons: 1) to avoid the identification of large and thematically inconsistent clusters, which could gather too distant knowledge, and 2) to allow a more precise partitioning of the dataset in the future, especially the big clusters on brain neural networks, neurological disorders ($C_7$), and central nervous system studies ($C_8$).

\subsection{Definition of the obtained clusters}
\label{app:definition_cluster}

Since the clusters delimited before are of different sizes, they could either cover a small and consistent lexical region corresponding to a given research specialty in neuroscience (such as $C_1$, $C_2$, $C_3$ and $C_4$), or many different ones and therefore many research areas with strong vocabulary distance at the same time (such as $C_7$ and $C_8$).
Furthermore, depending on the topical or conceptual taxonomy provided by bibliometric databases and used to label the articles in our dataset, we could misapprehend the contents of these clusters, notably the largest ones.
Therefore, in what follows, we compare three such taxonomies in order to approach as best as possible the contents of these clusters, here MAG's \textit{fields of study}, and OpenAlex's \textit{topics} and \textit{keywords}, which label the articles.
In particular, we indicate in Tab.~\ref{tab:MAG_OA_comp}, for each classification, the five most important elements shared by the articles in each cluster.

For the MAG classification, we complete the list of the most frequent fields of study situated at level 2 of the field network by indicating the most frequent ones at level 3, since the fields located at the former level are too general and do not provide a clear meaning of the cluster on their own.
Moreover, we provide this supplementary list because 92.3\% of the papers in our neuroscience dataset are tagged with at least one concept situated at this level. 
We discard the fields situated at the first level of the field networks because they are merely disciplines or broad research domain, and also the fourth and fifth ones because they are more sparsely tagged in our dataset.

\afterpage{%
\newgeometry{top=5mm, bottom=15mm}
\scriptsize
\newcommand*{\tabindent}{\hspace{3mm}}
\newcommand\Tstrut{\rule{0pt}{4ex}}         
\newcommand\Bstrut{\rule[-1.5ex]{0pt}{0pt}}   

\begin{landscape}
\setstretch{1}
\begin{longtable}{|l|l|l|l|l|}

\hline
\multicolumn{1}{|c}{} & \multicolumn{1}{|c|}{\normalsize{\textbf{Fields of Study (level 2)}}} & \multicolumn{1}{c|}{\normalsize{\textbf{Fields of Study (level 3)}}} & \multicolumn{1}{c|}{\normalsize{\textbf{Topics}}} & \multicolumn{1}{c|}{\normalsize{\textbf{Keywords}}} \Tstrut\Bstrut \\
\hline
\endfirsthead
\hline
\multicolumn{1}{|c}{} & \multicolumn{1}{|c|}{\normalsize{\textbf{Fields of Study (level 2)}}} & \multicolumn{1}{c|}{\normalsize{\textbf{Fields of Study (level 3)}}} & \multicolumn{1}{c|}{\normalsize{\textbf{Topics}}} & \multicolumn{1}{c|}{\normalsize{\textbf{Keywords}}} \Tstrut\Bstrut \\
\hline
\endhead
\hline
\addlinespace 
\multicolumn{5}{r}{\small{\textit{Continued on next page}}} \\
\endfoot
\hline
\addlinespace 
\caption{Table summarizing the five most important fields of study (retrieved from MAG classification, at levels 2 and 3), topics and keywords (both retrieved from OpenAlex) per cluster.
To fill the second column, we retain only OpenAlex's \textit{primary topics} of the papers within each cluster.
In the third column, we consider the ensemble of all OpenAlex's keywords associated with the articles within each cluster.}
\endlastfoot
                                            
\multirow{9}{*}{\large{$C_0$}} & Diabetes mellitus & Retinal & \begin{tabular}[c]{@{}l@{}}Molecular Mechanisms of \\ \tabindent Retinal Degeneration \\ \tabindent and Regeneration\end{tabular} & Intraocular Pressure  \\
& Population & Retinal  & \begin{tabular}[c]{@{}l@{}}Global Prevalence and \\ \tabindent Treatment of Glaucoma\end{tabular}  & Photoreceptor Degeneration \\
& In vivo  & Eye Disease & \begin{tabular}[c]{@{}l@{}}Age-Related Macular \\ \tabindent Degeneration Research\end{tabular}  & Intraocular Lens Implantation \\
& Cell culture  & Visual acuity  & \begin{tabular}[c]{@{}l@{}}Biomechanical Properties of \\ \tabindent the Cornea and Related Diseases\end{tabular} & \begin{tabular}[c]{@{}l@{}}Sutureless Intraocular \\ \tabindent Lens Fixation\end{tabular} \\
& Erg  & Cornea  & \begin{tabular}[c]{@{}l@{}}Cataract Surgery Techniques\\ \tabindent and Complications\end{tabular}  & Ocular Hypertension                                                             \\ 
\hline

\multirow{9}{*}{\large{$C_1$}} & Artificial neural network & Exponential stability                         & \begin{tabular}[c]{@{}l@{}}Neural Network Fundamentals\\ \tabindent  and Applications\end{tabular}                               & Backpropagation Learning                                                        \\
 & Nonlinear system  & Recurrent Neural Network                      & \begin{tabular}[c]{@{}l@{}}Network Synchronization in\\ \tabindent Complex Systems\end{tabular}                                 & Time Delays                                                                     \\
 & Support vector machine    & Lyapunov function                             & \begin{tabular}[c]{@{}l@{}}Face Recognition and\\ \tabindent Dimensionality Reduction Techniques\end{tabular}                   & Support Vector Machines                                                         \\
 & Cluster Analysis                       & Deep learning                                 & \begin{tabular}[c]{@{}l@{}}Theory and Applications\\ \tabindent of Extreme Learning Machines\end{tabular}                       & Feedforward Neural Networks                                                     \\
 & Convergence (routing)                                  & Backpropagation                               & \begin{tabular}[c]{@{}l@{}}Blind Source Separation and\\ \tabindent Independent Component Analysis\end{tabular}                 & Recurrent Neural Networks                                                       \\ \hline

\multirow{9}{*}{\large{$C_2$}} & \begin{tabular}[c]{@{}l@{}}Attention deficit \\ \tabindent hyperactivity disorder\end{tabular}  & Methylphenidate                               & \begin{tabular}[c]{@{}l@{}}Attention-Deficit/\\ \tabindent Hyperactivity Disorder\end{tabular}                                                                             & \begin{tabular}[c]{@{}l@{}}Attention-Deficit/\\ \tabindent Hyperactivity Disorder\end{tabular} \\
 & Cognition                                           & Comorbidity                                   & \begin{tabular}[c]{@{}l@{}}Analysis of Brain\\ \tabindent Functional Connectivity Networks\end{tabular}                         & ADHD                                                                            \\
   & Electroencephalography  & Neuropsychology                               & Autism Spectrum Disorders                                                                                            & Adolescent Brain Development                                                    \\
  &  Impulsivity                                        & El Niño                                       & \begin{tabular}[c]{@{}l@{}}Obsessive-Compulsive Disorder and\\ \tabindent Related Conditions\end{tabular}                       & Brain Imaging                                                                   \\
  & Population                                          & Working memory                                & \begin{tabular}[c]{@{}l@{}}Epidemiology and Management\\ \tabindent of Bipolar Disorder\end{tabular}                            & Attentional Networks                                                            \\ \hline

\multirow{9}{*}{\large{$C_3$}}  & Obsessive compulsive                  & Tourette syndrome                             & \begin{tabular}[c]{@{}l@{}}Obsessive-Compulsive Disorder\\ \tabindent and Related Conditions\end{tabular}                       & Obsessive-Compulsive Disorder                                                   \\
  & Tics                                          & Anxiety                                       & \begin{tabular}[c]{@{}l@{}}Deep Brain Stimulation\\ \tabindent for Neurological Disorders\end{tabular}                          & Tourette Syndrome                                                               \\
    & Cognition                                        & Comorbidity                                   & \begin{tabular}[c]{@{}l@{}}Analysis of Brain Functional\\ \tabindent Connectivity Networks\end{tabular}                         & Deep Brain Stimulation                                                          \\
       & Neuroimaging                                  & Schizophrenia                                 & \begin{tabular}[c]{@{}l@{}}Effects of Brain Stimulation \\ \tabindent on Motor Cortex\end{tabular}                              & Hoarding Behavior                                                               \\
 & Obsessive-compulsive disorder (OCD) & Neuropsychology                               & Diffusion Magnetic Resonance Imaging                                                                                 & Treatment                                                                       \\ \hline

\multirow{5}{*}{\large{$C_4$}}   & Gene                 & Disease                                       & Pathophysiology of Parkinson's Disease                                                                               & Parkinson's Disease                                                             \\
  & Population                                          & Mutation                                      & \begin{tabular}[c]{@{}l@{}}Lysosomal Storage Disorders \\ \tabindent in Human Health and Disease\end{tabular}                   & Parkinsonism                                                                    \\
  & Diabetes Mellitus                                          & Genotype                                      & \begin{tabular}[c]{@{}l@{}}Epigenetic Modifications and \\ \tabindent Their Functional Implications\end{tabular}               & Neurodegeneration                                                               \\
    & Meta-analysis                                     & Allele                                        & \begin{tabular}[c]{@{}l@{}}Nurr1/CoREST Pathway in \\ \tabindent Neuroprotection and Inflammation\end{tabular}                  & Dopaminergic Neurons                                                            \\
    & Cognition                                        & Polymorphism (computer science)               & \begin{tabular}[c]{@{}l@{}}Deep Brain Stimulation \\ \tabindent for Neurological Disorders\end{tabular}                         & Genetic Risk Factors                                                            \\ \hline

\multirow{9}{*}{\large{$C_5$}}  & Gene                   & Cancer                                        & Gliomas                                                                                                              & Skeletal Muscle Atrophy                                                         \\
  & Signal transduction                                          & Apoptosis                                     & \begin{tabular}[c]{@{}l@{}}Molecular Mechanisms of \\ \tabindent Muscle Regeneration and Atrophy\end{tabular}                   & Glioblastoma                                                                    \\
 & Diabetes mellitus                                           & Transcription factor                          & \begin{tabular}[c]{@{}l@{}}Regulation and Function of \\ \tabindent Microtubules in Cell Division\end{tabular}                  & mRNA modification                                                               \\
  & Population                                          & Receptor                                      & \begin{tabular}[c]{@{}l@{}}Mitochondrial Dynamics and \\ \tabindent Reactive Oxygen Species Regulation\end{tabular}             & Muscle Regeneration                                                             \\
    & Cell                                        & Mutation                                      & \begin{tabular}[c]{@{}l@{}}Myasthenia Gravis and \\ \tabindent Thymic Tumors Research\end{tabular}                              & Brain Tumor Epidemiology                                                        \\ \hline

\multirow{9}{*}{\large{$C_6$}} & Stimulus (physiology)                   & Electrophysiology                             & \begin{tabular}[c]{@{}l@{}}Cochlear Neuropathy and \\ \tabindent Hearing Loss Mechanisms\end{tabular}                           & Olfactory Receptors                                                             \\
  & Cochlea                                          & Hair cell                                     & \begin{tabular}[c]{@{}l@{}}Olfactory Dysfunction in \\ \tabindent Health and Disease\end{tabular}                               & Olfactory System                                                                \\
   & Olfaction                                         & Auditory cortex                               & \begin{tabular}[c]{@{}l@{}}Neuroscience and Genetics of \\ \tabindent Drosophila Melanogaster\end{tabular}                      & Auditory Processing                                                             \\
 & Inner ear                                           & Stimulation                                   & \begin{tabular}[c]{@{}l@{}}Avian Vocal Communication \\ \tabindent and Evolutionary Implications\end{tabular}                   & Inner Ear Development                                                           \\
  & Sensory system                                          & Receptor                                      & \begin{tabular}[c]{@{}l@{}}Impact of Hearing Loss \\ \tabindent on Cognitive Function\end{tabular}                              & Avian Vocal Communication                                                       \\ \hline

\multirow{9}{*}{\large{$C_7$}}  & Cognition                    & Schizophrenia                                 & \begin{tabular}[c]{@{}l@{}}Neural Mechanisms of \\ \tabindent Visual Perception and Processing\end{tabular}                     & Visual Perception                                                               \\
 & Stimulus (physiology)                                   & Epilepsy                                      & \begin{tabular}[c]{@{}l@{}}Neural Mechanisms of Cognitive \\ \tabindent Control and Decision Making\end{tabular}                & Neuroimaging Data Analysis                                                      \\
  & Perception                                          & Visual perception                             & Epilepsy and Seizures                                                                                                & Perceptual Learning                                                             \\
     & Electroencephalography                                       & Prefrontal cortex                             & \begin{tabular}[c]{@{}l@{}}Analysis of Brain Functional \\ \tabindent Connectivity Networks\end{tabular}                        & Working Memory                                                                  \\
   & Population                                         & Disease                                       & Neuronal Oscillations in Cortical Networks                                & Sensory Processing                                                              \\ \hline

\multirow{9}{*}{\large{$C_8$}} & Hippocampus                   & Central nervous system                        & \begin{tabular}[c]{@{}l@{}}Molecular Mechanisms of \\ \tabindent Synaptic Plasticity and \\ Neurological Disorders\end{tabular} & Neuroinflammation                                                               \\
   & Diabetes mellitus                                         & Receptor                                      & \begin{tabular}[c]{@{}l@{}}Neurobiological Mechanisms of \\ \tabindent Drug Addiction and Depression\end{tabular}               & Glutamate Receptors                                                             \\
    & Glutamate receptor                                        & Hippocampal formation                         & \begin{tabular}[c]{@{}l@{}}Role of Neuropeptides in \\ \tabindent Physiology and Disease\end{tabular}                           & Dopamine                                                                        \\
   & Spinal cord                                        & Dopamine                                      & Mechanisms of Alzheimer's Disease                                                                                     & Neurodegeneration                                                               \\
   & Neuron                                         & Stimulation                                   & \begin{tabular}[c]{@{}l@{}}Mechanisms and Management \\ \tabindent of Neuropathic Pain\end{tabular}                             & Cell Signaling                                                                 

\label{tab:MAG_OA_comp}

\end{longtable}
\end{landscape}
}

\restoregeometry 

Despite the excessive generality of the MAG taxonomy of fields of study -- even at the third level -- we observe that the three classifications match quite well for the smallest clusters: $C_0$\footnote{A closer look at the following level-2 fields of study within the $C_0$ ranking reveals the terms \textit{Visual field}, \textit{Lens (optics)} and \textit{Astigmatism}, thus confirming the name of the cluster mentioned in the main text.} 
focuses on the eye and vision studies, $C_1$ on the design of artificial neural network models and classifiers for various applications, $C_2$ on the Attention-Deficit/Hyperactivity Disorders (ADHD) -- the first term \textit{Methylphenidate} within the list of level-3 fields of study for this cluster is also a molecule found in psychostimulant treatments for such disorders --, $C_3$ on the neurological studies of the Obsessive-Compulsive Disorder (OCD) and Tourette syndrome, $C_4$ on the studies of the genetic factors responsible for the onset of Parkinson's disease, and $C_6$ on the studies of sensory pathologies related to hearing and smell.
The other three clusters are more thematically diversified: $C_5$ gathers studies on glioma, an aggressive brain tumor responsible for a severe neurological degeneration, and on skeletal muscle atrophy and regeneration, $C_7$ focuses on the studies of various functions of brain neural networks -- visionl, hearing, memory, cognition, etc. -- and also on shizophrenia and epilepsy disorders, and $C_8$\footnote{The following terms within the ranking of level-2 fields of study describing cluster $C_8$ are \textit{Cerebral cortex} and \textit{Synapse}, thus confirming the chosen name of the cluster centered on various studies of the central nervous system.} on various neurobiological mechanisms that could occur in the central nervous system and that could explain some pathologies -- drug addiction, depression, neuropathic pain, among others -- and other neurological disorders,\footnote{The following website offers a classification of all the neurological disorders referenced: \url{https://www.merckmanuals.com/home/brain,-spinal-cord,-and-nerve-disorders} (visited on July, 14 2024).} such as Alzheimer's disease and multiple sclerosis.

By compiling such elements associated with their papers, we attribute a unique name to the clusters obtained above, as summarized in Tab.~\ref{tab:pop_clusters}.

\begin{table}[t!]
    \centering
    \begin{tabular}{|c|c|c|}
        \hline
        \textbf{Cluster} & \textbf{Name} & \bm{$N_p$} \\
        \hhline{|=|=|=|}
        $C_0$ & Eyes and vision studies & 56,644 \\
        \hline
        \multirow{2}{*}{$C_1$} & Mathematical and computational & \multirow{2}{*}{8,755} \\
        & foundations of connectionist AI &  \\
        \hline
        $C_2$ & Attention-Deficit/Hyperactivity Disorders & 2,222 \\
        \hline
        \multirow{2}{*}{$C_3$} & Obsessive-Compulsive Disorders and & \multirow{2}{*}{1,407} \\    
         & Tourette syndrome studies & \\
        \hline
        $C_4$ & Parkinson's disease & 1,179 \\
        \hline
        $C_5$ & Gliomas and skeletal muscle atrophy studies & 66,465 \\
        \hline
        $C_6$ & Studies of sensorial pathologies & 28,382 \\
        \hline
        \multirow{2}{*}{$C_7$} & Brain neural networks and neurological & \multirow{2}{*}{266,593} \\
         &  disorders studies (epilepsy and schizophrenia) & \\
        \hline
        \multirow{2}{*}{$C_8$} & Studies of neurobiological mechanisms & \multirow{2}{*}{370,814} \\
        & in the central nervous system &  \\
        \hhline{|=|=|=|}
        \textbf{Noise} & -- & 55,230 \\
        \hline
    \end{tabular}
    \caption{Final description of the clusters in Fig.~\ref{fig:map}.
    We also recall the number of papers $N_p$ within the clusters obtained with HDBSCAN.}
    \label{tab:pop_clusters}
\end{table}

\setcounter{figure}{0}
\section{Robustness of the concept network at levels 2 and 3, and the AI-related conceptual coreness}
\label{app:conceptNet_topo}

Whereas the original concept network of our dataset includes fields of study at all levels from 0 to 5 (according to the MAG classification), we have selected those located at levels 2 or 3 only. 
As already highlighted in Sect.~\ref{sec:build_concept_net}, we chose these because they provide a good balance of specificity: not too general (like level 0, which corresponds to big disciplinary area) and not too specialized (like level 5, which corresponds to keywords from titles and abstracts).
Although most of the papers within our corpus are labeled with these levels 2 and 3 (93\%), does this selection still conserve the main topological properties of the original, all-levels concept network?
Also, does it affect the coreness' trends of AI-related concepts provided in the main text in Sects.~\ref{sec:build_concept_net} and \ref{sec:intrication}?

To answer these questions, we performed some robustness checks to make sure the results are reliable, based on four time-aggregated concept networks built upon different level ranges: the all-levels ones ($G_{012345}$), levels 2 and 3 ($G_{23}$, which we used in the main text), levels 0 to 3 ($G_{0123}$) and levels 2 to 5 ($G_{2345}$).
Specifically, we compared these network in two ways.
First, we evaluated the degree of topological similarity between these networks with various indicators describing them, all shown in Fig.~\ref{fig:topo_desc_network}.
Second, we computed the temporal evolution of AI-related concepts' corenesses within these networks according to the method exposed in Sect.~\ref{sec:build_concept_net}.
All of these are shown in Fig.~\ref{fig:coreness_topo_conceptNet}.

\subsection{Topological analysis}

\begin{figure}[t!]
    \centering
    \begin{tabular}{|l|c|c|c|c|}
        \hline
        & $G_{012345}$ & $G_{123}$ & $G_{23}$ & $G_{2345}$ \\
        \hhline{|=|=|=|=|=|}
        Number of nodes $N_{nodes}$ & 116,150 & 57,810 & 57,325 & 115,648\\
        Ratio $N_{nodes}/N_{G_{012345}}$ & 1 & $\sim 0,5$ & 0,49 & $\sim 1$ \\
        Total weight $W$ & 41,648,392 & 21,599,142 & 8,020,962 & 20,662,927 \\
        Ratio $W_/W_{G_{012345}}$ & 1 & 0,52 & 0,19 & ~0,5 \\
        Clustering coefficient $c$ & 0,06 & 0,07 & 0,12 & 0,11 \\
        \hline
    \end{tabular}
    \includegraphics[width=0.7\textwidth]{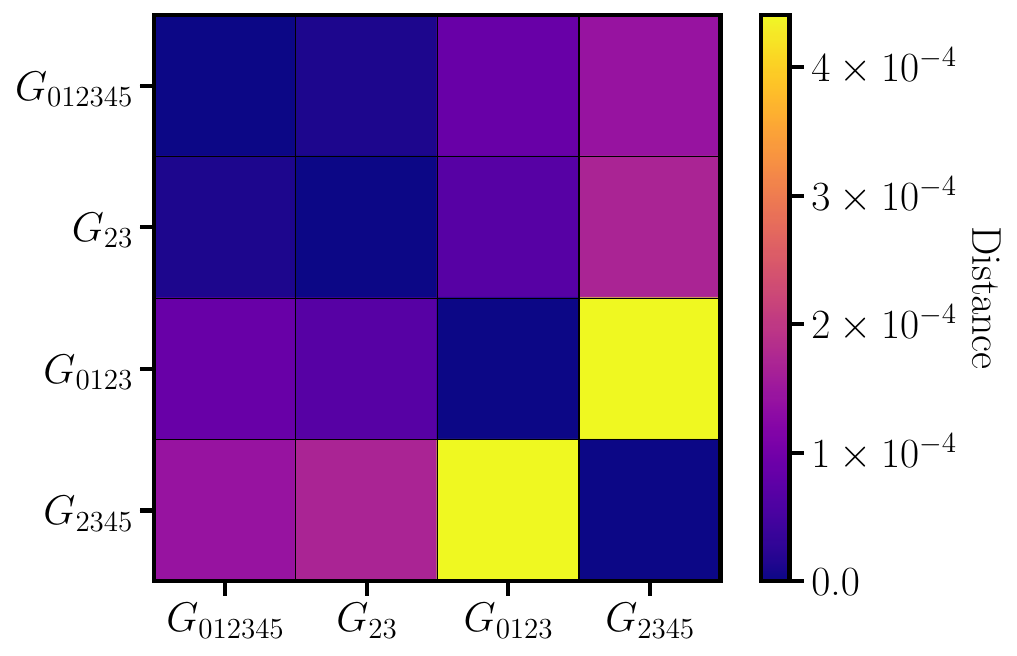}
    \caption{Summary of the robustness checks on the selection of fields of study situated at levels 2 and 3 within the concept network of our dataset $\mathcal{P}$.
    Top: Topological indicators describing the studied time-aggregated networks.
    The ratios of the number of nodes and total weights compared to those of the all-levels network $G_{012345}$ are also indicated to evaluate the loss of nodes and edge weight when selecting specific level ranges.
    Bottom: NetLSD distance matrix between these networks, computed with the method exposed in \citep{tsitsulin_netlsd_2018}.}
    \label{fig:topo_desc_network}
\end{figure}

Alongside the common metrics shown in the table in Fig.~\ref{fig:topo_desc_network}, we computed the Network Laplacian Spectral Distance matrix (NetLSD) with the algorithm proposed by Tsitsulin et al. (\citeyear{tsitsulin_netlsd_2018}) and that is available on GitHub.\footnote{\url{https://github.com/xgfs/NetLSD}}
In summary, this method consists in solving an information diffusion model on each network $G$, in which information quantity in each node at time $t$ is modeled by a vector $u_t$ that evolves according to the following heat equation:
\begin{equation}
    \frac{\partial u_t}{\partial t} = -\mathcal{L}u_t \, ,
\end{equation}
where $\mathcal{L}$ is the laplacian matrix of $G$, whose eigenvalues reflect the structural properties of $G$ at many scales.
The solution of such an equation is the temporal \textit{heat kernel} of the network, which is a sequence $H(G)=\{H_t\}_{t>0}$ of heat matrices between all the nodes $H_t=e^{-t\mathcal{L}}$ ($H_0$ is set by an initial heat $u_0$).
These heat matrices encode the quantity of information transferred time by time between two nodes and reflect the different centrality structures of the graph at all scales.
Tsitsulin et al. finally summarize such a diffusion process on $G$ with its scalar heat trace signature $h(G)=\{h_t\}_{t>0}$, where $h_t$ = diag$(H_t)$ is the trace of the associated heat matrix $H_t$ at time $t$.
To compare two network topologies, we rely on the cosine similarity of their respective heat signature $h$ encoding their topological information:
\begin{equation}
    d(G_1,G_2) = 1 - \frac{|h(G_1)\cdot h(G_2)|}{\|h(G_1)\|\|h(G_2)\|}
\end{equation}
This distance takes its values between 0 and 1.
If the distance is close to 0, then the two compared networks have been submitted to a rather common diffusion process between their respective nodes, which means that their respective topology are similar.

Although the table in Fig.~\ref{fig:topo_desc_network} exhibits a considerable differences between the network $G_{23}$ mobilized in the main text and all the others, especially in terms of conserved weight, we notice that the network that is topologically the closest to the all-levels one is effectively ours.
This result thus motivated us to retain $G_{23}$ to study the evolution of the location of AI-related concepts within the neuroscience concept network.

\subsection{AI coreness analysis}

\begin{figure}[t!]
    \centering
    \includegraphics[width=\textwidth]{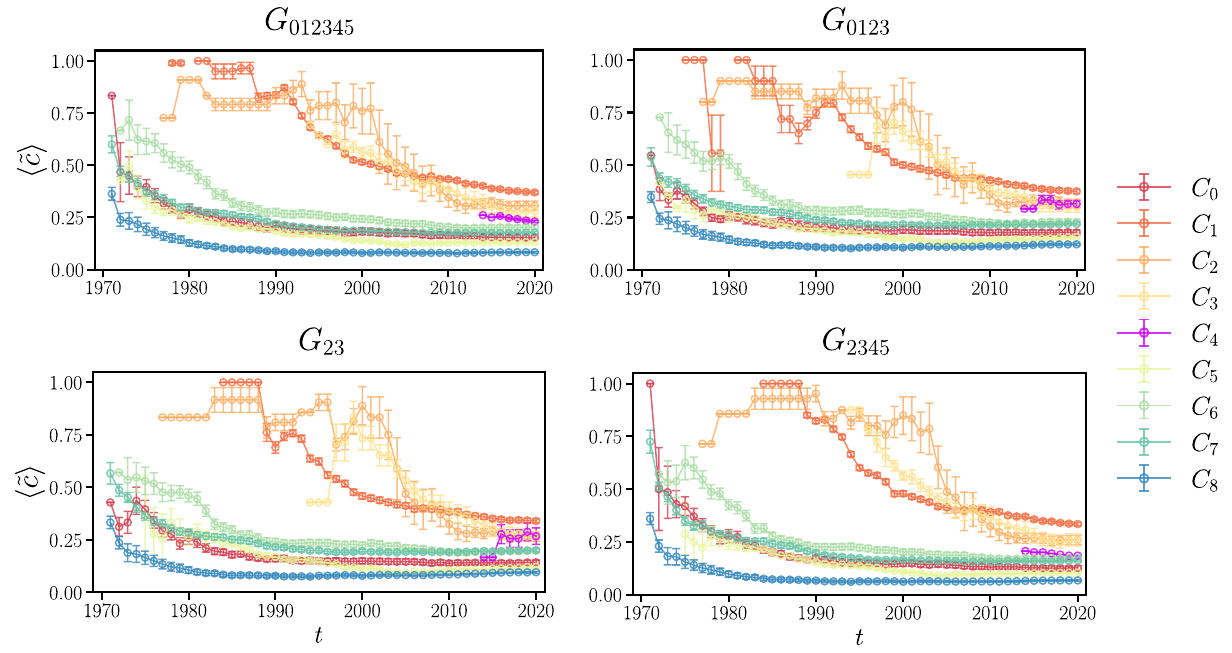}
    \caption{Temporal average coreness $\langle\tilde{c}\rangle$ of the AI-related concepts in the cumulative temporal concept networks of each cluster within each networks -- normalized by the highest coreness returned by the $k$-core decomposition of the network at a given year $t$.
    The higher the coreness $\langle\tilde{c}\rangle$, the closer the AI-related concepts are to the core of the network, and vice versa.
    We only consider AI-related concepts included in the giant component of each cluster's concept network.
    The error bars are the standard errors produced by the distribution of corenesses of all AI-related concepts within each cluster at each year, i.e. the ratio of the standard deviation to the square root of the number of entities present in the distribution.}
    \label{fig:coreness_topo_conceptNet}
\end{figure}

In this section, we consider the temporal versions of the previous concept networks at different levels.
According to the method provided in Sect.~\ref{sec:build_concept_net}, we computed the temporal average coreness of AI-related concepts within each cluster of the knowledge map (see Fig.~\ref{fig:map} in the main text).
As shown in Fig.~\ref{fig:coreness_topo_conceptNet}, the corenesses are very similar in every network, especially within the clusters on the studies of the formal foundations of connectionist AI ($C_1$), ADHD ($C_2$) and OCD ($C_3$).
Only values in the cluster of sensory pathology studies ($C_6$) and  between 1970 and early 1980s are lower in $G_{23}$ than in the three others, as well as those in the cluster of vision studies ($C_0$) between 1970 and 1975.
With the topological comparison provided in the previous section, this second result thus supports the choice of considering only the concept network with fields of study at levels 2 or 3.
Since it has fewer nodes and a lower total weight than $G_{012345}$, it thus allowed us to perform our analyses with less data and in less time, while keeping the main structures of the original one.

\setcounter{figure}{0}
\section{Dynamic citation radius of gyration of the papers published in 1976 and 1989}
\label{app:dynamic_gyration_radius}

The temporal average citation radius of gyration (RoG) produced by the location of the citations of a single paper within the neuroscience knowledge map, as shown in Fig.~\ref{fig:gyradius_citation_average} in the main text (under its normalized definition), is a measure that covers all the citations accumulated by that paper over the years.
Thus, a paper published at the very beginning of the time period under study (1970-2019) could have more citations and therefore a better lexical coverage than a paper published more recently.
In order to prospect the dynamic citation diffusion of the AI-related papers published in the two specific years when the average RoG exhibits a maximum, namely 1976 and 1989, we explore the evolution of the citation gyration radii of these papers, expressed here as the logarithmic return $\log(1+R_t)$, where $R_t=(r_g(t)-r_g(t-1))/r_g(t-1)$ is the percentage of gain ($R_t>0$) or loss ($R_t<0$) in RoG between two consecutive years.
This measure, shown in Fig.~\ref{fig:gyradius_citation_temporal} below, indicates for a single paper, respectively, the expansion of the lexical coverage exhibited by the papers citing it, or the concentration of these citing papers around the local lexical subspace surrounding the paper under study.

\begin{figure}[t!]
    \centering
    \includegraphics[width=\textwidth]{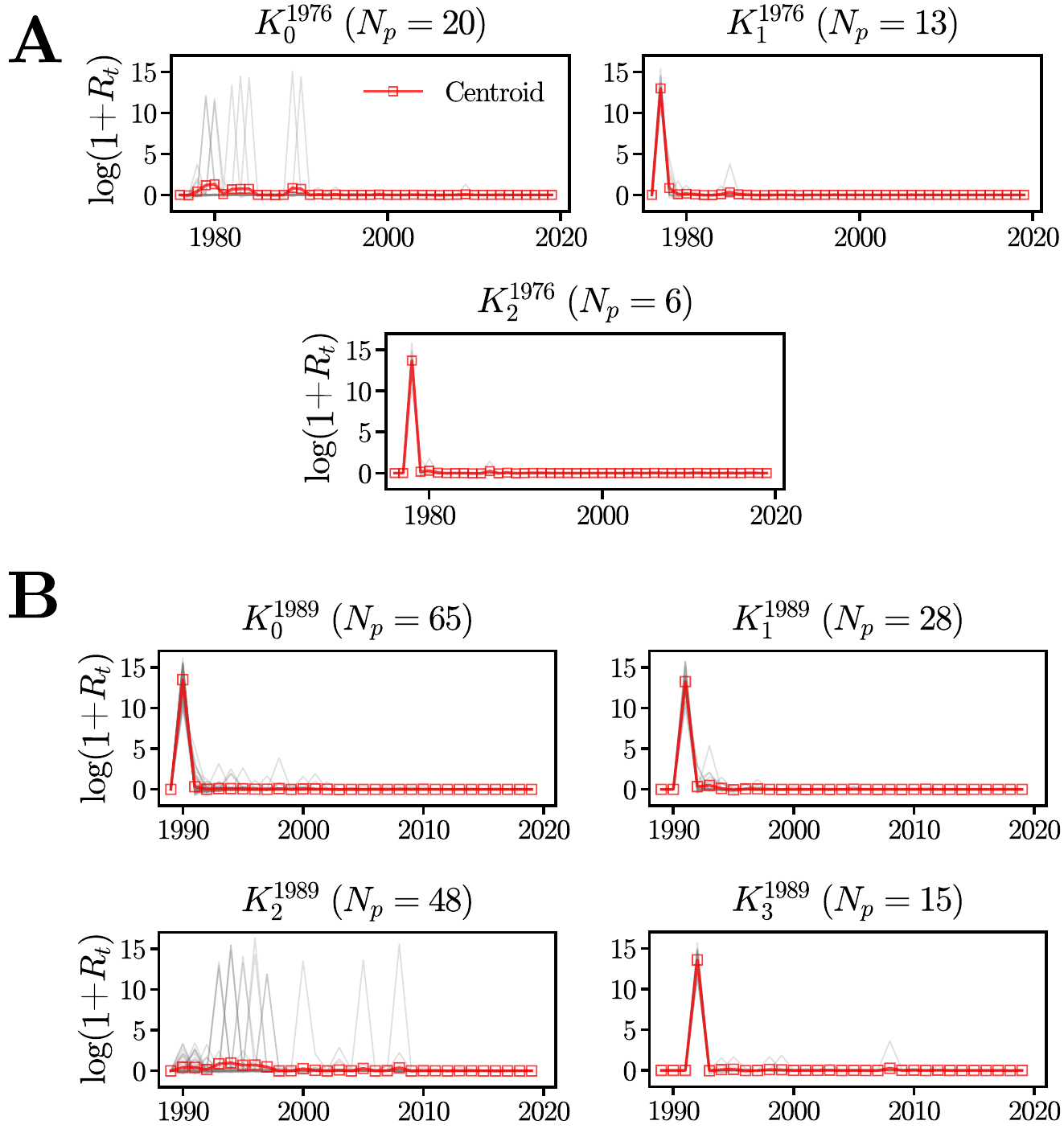}
    \caption{Temporal log return $R_t$ of the citation RoG of the papers published in 1976 (\textbf{A}) and in 1989 (\textbf{B}).
    The time series are aggregated into clusters $K_i$ in each year with a \textit{k-means} algorithm, based on an optimal number of clusters identified with a standard elbow method.
    $N_p$ is the number of articles within the studied cluster $K_i$ and therefore the number of different log-returned RoGs plotted for this cluster.
    The red curves represent the centroid of the cluster time series.}
    \label{fig:gyradius_citation_temporal}
\end{figure}

Fig.~\ref{fig:gyradius_citation_temporal} shows that most of the papers published in the two years studied (1976 and 1989) cover their final lexical subspace very quickly after the publication, as shown by the high peaks situated one, two or at most three years after publication.
This trend is represented by the papers included in the clusters $K^{1976}_1$, $K^{1976}_2$, $K^{1989}_0$, $K^{1989}_1$ and $K^{1989}_3$.
The papers included in the clusters $K^{1976}_0$ and $K^{1989}_2$ follow the same trend, but with some peculiarities, such as a variation that is not as important as those in the other clusters in the two or three years following the publication, and also a later impact after publication -- four years or more.

\end{appendices}
\clearpage

\bibliography{NeuroAI_topicModeling}

@misc{cohan_specter_2020,
	title = {{SPECTER}: {Document}-level {Representation} {Learning} using {Citation}-informed {Transformers}},
	note = "Preprint at \url{http://arxiv.org/abs/2004.07180}",
	author = {Cohan, Arman and Feldman, Sergey and Beltagy, Iz and Downey, Doug and Weld, Daniel S.},
	year = {2020},
}

@misc{beltagy_scibert_2019,
	title = {{SciBERT}: {A} {Pretrained} {Language} {Model} for {Scientific} {Text}},
	author = {Beltagy, Iz and Lo, Kyle and Cohan, Arman},
	year = {2019},
	note = "Preprint at \url{http://arxiv.org/abs/1903.10676}",
}

@misc{batagelj_om_2003,
	title = {An {O}(m) {Algorithm} for {Cores} {Decomposition} of {Networks}},
	author = {Batagelj, V. and Zaversnik, M.},
	year = {2003},
	note = "Preprint at \url{http://arxiv.org/abs/cs/0310049}",
}

@article{fontaine_epistemic_2024,
	title = {Epistemic integration and social segregation of {AI} in neuroscience},
	author = {Fontaine, Sylvain and Gargiulo, Floriana and Dubois, Michel and Tubaro, Paola},
	year = {2024},
    volume = {9},
	number = {8},
	journal = {Applied Network Science},
}

@misc{fontaine_epistemic_2024-1,
	title = {Epistemic integration and social segregation of {AI} in neuroscience - {Dataset} extracted from the {Microsoft} {Academic} {Knowledge} {Graph}},
	author = {Fontaine, Sylvain and Gargiulo, Floriana and Dubois, Michel and Tubaro, Paola},
    year = {2024},
	note = {Available on Zenodo at \url{https://zenodo.org/records/10777508}},
}

@misc{mcinnes_umap_2020,
	title = {{UMAP}: {Uniform} {Manifold} {Approximation} and {Projection} for {Dimension} {Reduction}},
	author = {McInnes, Leland and Healy, John and Melville, James},
	year = {2020},
	note = "Preprint at \url{http://arxiv.org/abs/1802.03426}",
}

@article{mcinnes_hdbscan_2017,
	title = {{HDBSCAN}: {Hierarchical} density based clustering},
	volume = {2},
	number = {11},
	journal = {The Journal of Open Source Software},
	author = {McInnes, Leland and Healy, John and Astels, Steve},
	year = {2017},
	pages = {205},
}

@article{petrovich_neuro-philo_2024,
	title = "Mapping the philosophy and neuroscience nexus through citation analysis",
	volume = "14",
	number = "60",
	journal = "European {Journal} for {Philosophy} of {Science}",
	author = "Petrovich, Eugenio and Viola, Marco",
	year = "2024"
}

@article{herfeld_role-typo_2019,
	title = "The diffusion of scientific innovations: {A} role typology",
	volume = "77",
	journal = "Studies in {History} and {Philosophy} of {Science} {Part} {A}",
	author = "Herfeld, Catherine and Doehne, Malte",
	year = "2019",
    pages = {64--80},
}

@article{di-bona_decentralization_2023,
	title = "The concept of decentralization through time and disciplines: a quantitative exploration",
	volume = "12",
    number = "1",
	journal = "{EPJ} {Data} {Science}",
	author = "Di Bona, Gabriele and Bracci, Alberto and Perra, Nicola and Latora, Vito and Baronchelli, Andrea",
	year = "2023",
    pages = "42",
}

@article{lariviere_aging_2009,
	title = "Long-term variations in the aging of scientific literature: {From} exponential growth to steady-state science (1900–2004)",
	volume = "59",
    number = "2",
	journal = "{Journal} of the {American} {Society} for {Information} {Science} and {Technology}",
	author = "Larivière, Vincent and Archambault, Eric and Gingras, Yves",
	year = "2008",
    pages = "288--296",
}

@incollection{ghidini_microsoft_2019,
	title = {The {Microsoft} {Academic} {Knowledge} {Graph}: {A} {Linked} {Data} {Source} with 8 {Billion} {Triples} of {Scholarly} {Data}},
	volume = {11779},
	booktitle = {The {Semantic} {Web} – {ISWC} 2019},
	publisher = {Springer International Publishing},
	author = {Färber, Michael},
	editor = {Ghidini, Chiara and Hartig, Olaf and Maleshkova, Maria and Svátek, Vojtěch and Cruz, Isabel and Hogan, Aidan and Song, Jie and Lefrançois, Maxime and Gandon, Fabien},
	year = {2019},
	pages = {113--129},
}

@article{monechi_waves_2017,
	title = {Waves of novelties in the expansion into the adjacent possible},
	volume = {12},
	number = {6},
	journal = {Plos One},
	author = {Monechi, Bernardo and Ruiz-Serrano, {\~A}lvaro and Tria, Francesca and Loreto, Vittorio},
	year = {2017},
}

@book{kauffman_investigations_2000,
	title = {Investigations},
	publisher = {Oxford University Press},
	author = {Kauffman, Stuart A.},
	year = {2000},
}

@article{frank_evolution_2019,
	title = {The evolution of citation graphs in artificial intelligence research},
	volume = {1},
	number = {2},
	journal = {Nature Machine Intelligence},
	author = {Frank, Morgan R. and Wang, Dashun and Cebrian, Manuel and Rahwan, Iyad},
	year = {2019},
	pages = {79--85},
}

@article{marcovich_science_2020,
	title = {Science research regimes as architectures of knowledge in context: {A} ‘longue durée’ comparative historical sociology of structures and dynamics in science},
	volume = {59},
	number = {2},
	journal = {Social Science Information},
	author = {Marcovich, Anne and Shinn, Terry},
	year = {2020},
	pages = {310--328}
}

@article{tsitsulin_netlsd_2018,
	title = {{NetLSD}: {Hearing} the {Shape} of a {Graph}},
	journal = {Proceedings of the 24th ACM SIGKDD International Conference on Knowledge Discovery \& Data Mining},
	author = {Tsitsulin, Anton and Mottin, Davide and Karras, Panagiotis and Bronstein, Alex and Müller, Emmanuel},
	year = {2018},
	pages = {2347--2356}
}

@misc{baruffaldi_identifying_2020,
	title = {Identifying and measuring developments in artificial intelligence: {Making} the impossible possible},
	author = {Baruffaldi, Stefano and van Beuzekom, Brigitte and Dernis, Hélène and Harhoff, Deitmar and Rao, Nandan and Rosenfeld, David and Squicciarini, Mariagrazia},
	year = {2020},
       note = {{OECD} {Science}, {Technology} and {Industry} {Working} {Papers}}
}

@article{gargiulo_meso-scale_2023,
	title = {A meso-scale cartography of the {AI} ecosystem},
    volume = {4},
	number = {3},
	journal = {Quantitative Science Studies},
	author = {Gargiulo, Floriana and Fontaine, Sylvain and Dubois, Michel and Tubaro, Paola},
	year = {2023},
	pages = {574--593}
}

@article{cardon_neurons_2018,
	title = {Neurons spike back. {The} invention of inductive machines and the artificial intelligence controversy},
	volume = {211},
	number = {5},
	journal = {R\'eseaux},
	author = {Cardon, Dominique and Cointet, Jean-Philippe and Mazières, Antoine},
	translator = {Carey-Libbrecht, Liz},
	year = {2018},
	pages = {173--220}
}

@article{hassabis_neuroscience-inspired_2017,
	title = {Neuroscience-{Inspired} {Artificial} {Intelligence}},
	volume = {95},
	number = {2},
	journal = {Neuron},
	author = {Hassabis, Demis and Kumaran, Dharshan and Summerfield, Christopher and Botvinick, Matthew},
	year = {2017},
	pages = {245--258}
}

@article{gopinath_artificial_2023,
	title = {Artificial intelligence and neuroscience: {An} update on fascinating relationships},
	volume = {125},
	journal = {Process Biochemistry},
	author = {Gopinath, Nishanth},
	year = {2023},
	pages = {113--120}
}

@article{perconti_deep_2020,
	title = {Deep learning and cognitive science},
	volume = {203},
	journal = {Cognition},
	author = {Perconti, Pietro and Plebe, Alessio},
	year = {2020},
	pages = {12}
}

@article{bianchini_artificial_2022,
	title = {Artificial intelligence in science: {An} emerging general method of invention},
	volume = {51},
	number = {10},
	journal = {Research Policy},
	author = {Bianchini, Stefano and Müller, Moritz and Pelletier, Pierre},
	year = {2022}
}

@misc{gao_quantifying_2023,
	title = "Quantifying the Benefit of Artificial Intelligence for Scientific Research",
	note = "Preprint at \url{https://arxiv.org/abs/2304.10578}",
	author = "Gao, Jian and Wang, Dashun",
	year = "2023"
}

@article{shinn_transverse_2002,
	title = {The {Transverse} {Science} and {Technology} {Culture}: {Dynamics} and {Roles} of {Research}-{Technology}},
	volume = {41},
	number = {2},
	journal = {Social Science Information},
	author = {Shinn, Terry and Joerges, Bernward},
	year = {2002},
	pages = {207--251}
}

@article{hentschel_periodization_2015,
	title = {A periodization of research technologies and of the emergency of genericity},
	volume = {52},
	journal = {Studies in History and Philosophy of Modern Physics},
	author = {Hentschel, Klaus},
	year = {2015},
	pages = {223--233}
}

@article{xu_artificial_2021,
	title = {Artificial intelligence: {A} powerful paradigm for scientific research},
	volume = {2},
	number = {4},
	journal = {The Innovation},
	author = {Xu, Yongjun and Liu, Xin and Cao, Xin and Huang, Changping and Liu, Enke and Qian, Sen and Liu, Xingchen and Wu, Yanjun and Dong, Fengliang and Qiu, Cheng-Wei and Qiu, Junjun and Hua, Keqin and Su, Wentao and Wu, Jian and Xu, Huiyu and Han, Yong and Fu, Chenguang and Yin, Zhigang and Liu, Miao and Roepman, Ronald and Dietmann, Sabine and Virta, Marko and Kengara, Fredrick and Zhang, Ze and Zhang, Lifu and Zhao, Taolan and Dai, Ji and Yang, Jialiang and Lan, Liang and Luo, Ming and Liu, Zhaofeng and An, Tao and Zhang, Bin and He, Xiao and Cong, Shan and Liu, Xiaohong and Zhang, Wei and Lewis, James P. and Tiedje, James M. and Wang, Qi and An, Zhulin and Wang, Fei and Zhang, Libo and Huang, Tao and Lu, Chuan and Cai, Zhipeng and Wang, Fang and Zhang, Jiabao},
	year = {2021},
	pages = {100179},
}

@article{rumelhart_learning_1986,
	title = {Learning representations by back-propagating errors},
	volume = {323},
    number = {6088},
	journal = {Nature},
	author = {Rumelhart, David E. and Hinton, Geoffrey E. and Williams, Ronald J.},
	year = {1966},
	pages = {533-536},
}

@article{nmi_neuroAI_2024,
	title = {The new {NeuroAI}},
	volume = {6},
    number = {3},
	journal = {Nature Machine Intelligence},
	author = {},
	year = {2024},
	pages = {245-245},
}

@article{bornmann_citations_2008,
	title = {What do citation counts measure? {A} review of studies on citing behavior},
	volume = {64},
    number = {1},
	journal = {Journal of {Documentation}},
	author = {Bornmann, Lutz and Daniel, Hans-Dieter},
	year = {2008},
	pages = {45--80},
}

@article{ahmed_growing_2023,
	title = {The growing influence of industry in {AI} research},
	volume = {379},
    number = {6635},
	journal = {Science},
	author = {Ahmed, Nur and Wahed, Muntasir and Thompson, Neil C.},
	year = {2023},
	pages = {884-886},
}

@article{anichini_radio_2021,
	title = {L’intelligence artificielle à l’épreuve des savoirs tacites : analyse des pratiques d’utilisation d’un outil d’aide a la détection en radiologie},
	volume = {39},
    number = {2},
	journal = {Sciences Sociales et Santé},
	author = {Anichini, Giulia and Geffroy, Bénédicte},
	year = {2021},
	pages = {43-69},
}

@article{mignot_radio_2022,
	title = {Les innovations d’intelligence artificielle en radiologie à l’épreuve des régulations du système de santé},
	volume = {232-233},
    number = {2},
	journal = {Réseaux},
	author = {Mignot, Léo and Schultz, \'Emilien},
	year = {2022},
	pages = {65-97},
}

@misc{hajkowicz_csiro_2022,
	title = {{Artificial} intelligence for science -- {Adoption} trends and future development pathways},
	author = {Hajkowicz, S. and Naughtin, C. and Sanderson, C. and Schleiger, E. and Karimi, S. and Bratanova, A. and Bednarz, T.},
	year = {2022},
    note = {{CSIRO} {Data61}, {Brisbane}, {Australia}}
}

@article{fregnac_neuro_2017,
	title = {Big data and the industrialization of neuroscience: {A} safe roadmap for understanding the brain?},
	volume = {358},
    number = {6362},
	journal = {Science},
	author = {Frégnac, Yves},
	year = {2017},
	pages = {470-477},
}

@misc{klinger_narrowing_2022,
	title = {A narrowing of {AI} research?},
	language = {en},
	publisher = {arXiv},
	author = {Klinger, Joel and Mateos-Garcia, Juan and Stathoulopoulos, Konstantinos},
	year = {2022},
	note = {arXiv:2009.10385v4},
}

@misc{ahmed_dedemo_2020,
	title = {The {De}-democratization of {AI}: {Deep} {Learning} and the {Compute} {Divide} in {Artificial} {Intelligence} {Research}},
	language = {en},
	publisher = {arXiv},
	author = {Ahmed, Nur and Wahed, Muntasir},
	year = {2020},
	note = {arXiv:2010.15581},
}

@article{singh_charting_2024,
	title = {Charting mobility patterns in the scientific knowledge landscape},
	volume = {13},
	number = {1},
	journal = {EPJ Data Science},
	author = {Singh, Chakresh Kumar and Tupikina, Liubov and Lécuyer, Fabrice and Starnini, Michele and Santolini, Marc},
	year = {2024},
	pages = {1--20},
}

@article{liu_science_2024,
	title = {Science as exploration in a knowledge landscape: tracing hotspots or seeking opportunity?},
	volume = {13},
	number = {1},
	journal = {EPJ Data Science},
	author = {Liu, Feifan and Zhang, Shuang and Xia, Haoxiang},
	year = {2024},
	pages = {1--20},
}

@article{gonzalez-marquez_landscape_2024,
	title = {The landscape of biomedical research},
	volume = {5},
	number = {6},
	journal = {Patterns},
	author = {González-Márquez, Rita and Schmidt, Luca and Schmidt, Benjamin M. and Berens, Philipp and Kobak, Dmitry},
	year = {2024},
	pages = {100968},
}

@article{rosenblatt_perceptron_1958,
	title = {The perceptron: {A} probabilistic model for information storage and organization in the brain},
	volume = {65},
	number = {6},
	journal = {Psychological Review},
	author = {Rosenblatt, Frank},
	year = {1958},
	pages = {386--408},
}

@article{macpherson_natural_2021,
	title = {Natural and {Artificial} {Intelligence}: {A} brief introduction to the interplay between {AI} and neuroscience research},
	volume = {144},
	journal = {Neural Networks},
	author = {Macpherson, Tom and Churchland, Anne and Sejnowski, Terry and DiCarlo, James and Kamitani, Yukiyasu and Takahashi, Hidehiko and Hikida, Takatoshi},
	year = {2021},
	pages = {603--613},
}

@article{marcovich_regimes_2012,
	title = {Regimes of science production and diffusion: towards a transverse organization of knowledge},
	volume = {10},
	number = {spe},
	journal = {Scientiae Studia},
	author = {Marcovich, Anne and Shinn, Terry},
	year = {2012},
	pages = {33--64},
}

@misc{cockburn_impact_2018,
	title = {The {Impact} of {Artificial} {Intelligence} on {Innovation}},
	note = {National Bureau of Economic Research},
	author = {Cockburn, Iain and Henderson, Rebecca and Stern, Scott},
	year = {2018},
}

@article{roth_social_2010,
	title = {Social and semantic coevolution in knowledge networks},
	number = {1},
	journal = {Social Networks},
	author = {Roth, Camille and Cointet, Jean-Philippe},
	year = {2010},
	pages = {16--29},
}

@article{wray_rethinking_2005,
	title = {Rethinking {Scientific} {Specialization}},
	volume = {35},
	number = {1},
	journal = {Social Studies of Science},
	author = {Wray, K. Brad},
	year = {2005},
	pages = {151--164},
}

@article{rule_lexical_2015,
	title = {Lexical shifts, substantive changes, and continuity in {State} of the {Union} discourse, 1790–2014},
	volume = {112},
	number = {35},
	journal = {Proceedings of the National Academy of Sciences},
	author = {Rule, Alix and Cointet, Jean-Philippe and Bearman, Peter S.},
	year = {2015},
	pages = {10837--10844},
}

@misc{brahim_data-driven_2021,
	title = {A data-driven analysis to question epidemic models for citation cascades on the blogosphere},
	note = "Preprint at \url{http://arxiv.org/abs/1306.0424}",
	author = {Brahim, Abdelhamid Salah and Tabourier, Lionel and Le Grand, Bénédicte},
	year = {2021},
}

@article{huang_number_2018,
	title = {Number versus structure: towards citing cascades},
	volume = {117},
	number = {3},
	journal = {Scientometrics},
	author = {Huang, Yong and Bu, Yi and Ding, Ying and Lu, Wei},
	year = {2018},
	pages = {2177--2193},
}

@article{cheng_how_2023,
	title = {How {New} {Ideas} {Diffuse} in {Science}},
	volume = {88},
	number = {3},
	journal = {American Sociological Review},
	author = {Cheng, Mengjie and Smith, Daniel Scott and Ren, Xiang and Cao, Hancheng and Smith, Sanne and McFarland, Daniel A.},
	year = {2023},
	pages = {522--561},
}

\end{document}